\documentclass[aps,jmp,amsmath,amssymb,reprint]{revtex4-1}
\usepackage{graphicx}% Include figure files
\usepackage{dcolumn}% Align table columns on decimal point
\usepackage{bm}% bold math
\usepackage{graphicx}
\usepackage{subfigure}
\usepackage{physics}
\usepackage{color}
\begin{document}

%\preprint{AIP/123-QED}
\title[Physical Review]{Superlattice design for optimal thermoelectric generator performance}

\author{Pankaj Priyadarshi}
\author{Abhishek Sharma}
\author{Swarnadip Mukherjee}
\author{Bhaskaran Muralidharan}%
\email{bm@ee.iitb.ac.in}
\affiliation{Department of Electrical Engineering, Indian Institute of Technology Bombay, Powai, Mumbai 400076, India}

% \homepage{http://www.Second.institution.edu/~Charlie.Author.}
%\affiliation{%
%Second institution and/or address%\\This line break forced% with \\
%}%


\begin{abstract}
We consider the design of an optimal superlattice thermoelectric generator via the energy bandpass filter approach. Various configurations of superlattice structures are explored to obtain a bandpass transmission spectrum that approaches the ideal ``boxcar'' form, which is now well known to manifest the largest efficiency at a given output power.  Using the non-equilibrium Green's function formalism coupled self-consistently with the Poisson's equation,  we identify such an ideal structure and also demonstrate that it is almost immune to the deleterious effect of self-consistent charging and device variability. Analyzing various superlattice designs, we conclude that superlattices with a Gaussian distribution of the barrier thickness offers the best thermoelectric efficiency at maximum power. It is observed that the best operating regime of this device design provides a maximum power in the range of 0.32-0.46 $MW/m^2$ at efficiencies between 54\%-43\% of Carnot efficiency. We also analyze our device designs with the conventional figure of merit approach to counter support the results so obtained. We note a high $zT_{el}=6$ value in the case of Gaussian distribution of the barrier thickness. With the existing advanced thin-film growth technology, the suggested superlattice structures can be achieved, and such optimized thermoelectric performances can be realized.
\end{abstract}

\maketitle

\section{Introduction}
\begin{figure}
	\centering
	\includegraphics[height=0.7\textwidth,width=0.48\textwidth]{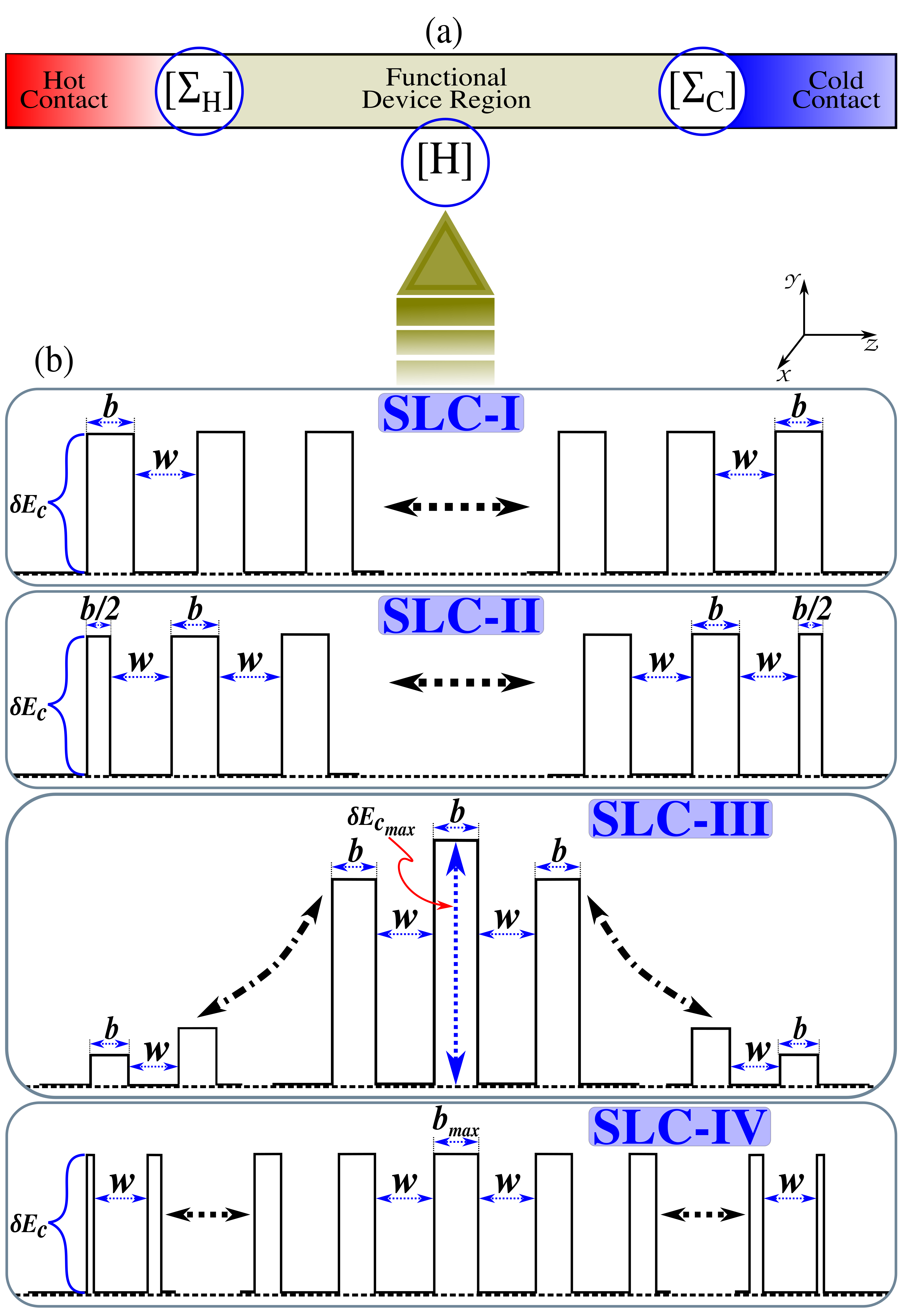}
		\caption{(a) A quantum transport treatment of a typical thermoelectric generator set up entails that the central region of the device is described by the Hamiltonian $[H]$ which is connected via the self-energy matrices $[\Sigma_{H, C}]$ to the hot and cold contacts. (b) The different superlattice configurations (SLCs) that form the central region, where $w$ is the width of the well region, $b$ is the barrier thickness, and $\delta E_c$ is the barrier height. SLC-I is the regular SL structure having a constant well width and barrier thickness. SLC-II is the AR-enabled SL, in which two additional barriers of half the thickness of a regular barrier is attached after a well width at both the ends. SLC-III and SLC-IV represent the Gaussian distribution of barrier height and barrier thickness respectively.}
	\label{Device}
\end{figure}

The field of nanostructured thermoelectrics (TE) \cite{Hicks1992, Hicks1993, Hicks1996, Heremans2013} is now well established for the prospect of achieving high conversion efficiencies in contrast with their bulk counterparts \cite{Majumdar2004, Snyder2008, Aniket2015, Mao2016}. Conventionally, a dimensionless figure of merit $zT=S^2 \sigma / (\kappa_{el}+\kappa_{ph})$, is employed in order to gauge the efficiency of a thermoelectric material, where $S$ is the Seebeck coefficient (thermopower), $\sigma$ is the electric conductivity, $\kappa_{el}$ and $\kappa_{ph}$ are the electronic and the lattice contributions to the thermal conductivity, and $T$ is the operating temperature. However, recent studies from a nanoscale transport theory perspective \cite{Nakpathomkun2010, Muralidharan2012, Sothmann2013, Agarwal2014} have shown that a high $zT$ does not necessarily ensure a functional thermoelectric generator in terms of the actual delivered power output. An important instance of which is that while it has been mathematically proposed \cite{Sofo-Mahan1996} that a Dirac delta transmission function ensures a conversion efficiency at the Carnot limit,  $\eta_C = 1-T_C/T_H$, where $T_H$ and $T_C$ are the temperatures at the hot and cold contacts respectively (Fig. \ref{Device}), the output power of such a device is zero \cite{Humphrey2005}. In the parlance of the figure of merit, such a structure possesses an infinite value of $zT$, and actually delivers a zero power output. This typically establishes the trade-off between the efficiency and the output power \cite{Muralidharan2012, Barajas-Aguilar2013, Bitan2016} in nanostructured thermoelectric devices, and also the inadequacy of using the figure of merit as the sole performance descriptor. \\
\indent In order to reconcile with this issue, it was recently proposed by Whitney \cite{Whitney2014,Whitney2015} that a ``boxcar" shaped transmission function with a finite spectral width can serve to maximize the efficiency of an irreversible system for a given output power. The boxcar shaped transmission function can in general be achieved by using heterostructure layers, such as a superlattice (SL), in which minibands are formed with a certain bandwidth separated by forbidden bands \cite{Tung1996, Pacher2001, Morozov2002, Gomez1999, Barajas-Aguilar2013} due to which a favorable density of states (DOS) profile for providing a thermoelectric figure of merit enhancement \cite{Sofo-Mahan1994, Broido1995, Balandin2003} results. Furthermore, superlattices can be designed for optimizing electronic transport by controlling the band offsets, quantum confinements and the tunneling processes between the different layers of materials \cite{Tsu2010}. Hence the design of appropriate superlattices with the objective of optimizing thermoelectric performance is of current and imminent interest. \\
\indent Based on the above findings, a recent work by Karbaschi et al., \cite{Karbaschi2016} detailed a power-efficiency analysis of a thermoelectric nanowire set up comprising of an anti-reflection enabled superlattice, which can potentially engineer a rectangular boxcar shaped transmission. However, the superlattice structures proposed there have strong multiple lineshape imperfections that reduces the transmissivity (area under the transmission curve) considerably, which becomes more drastic when one considers charging effects. In order to overcome this problem, in this paper, we incorporate and investigate in detail, the various strategies to achieve a robust boxcar transmission which is almost immune to such realistic charging effects. Furthermore, we provide a comparative study of various SL configurations to single out the structure that manifests the highest achievable efficiency at a given output power. \\
\indent Analyzing a voltage controlled SL-TE generator set up theoretically \cite{Nakpathomkun2010, Muralidharan2012, Sothmann2013, Agarwal2014, Aniket2015}, we investigate the performance in terms of output power and efficiency from an electrical engineering perspective \cite{Zebarjadi2012, Saniya2014} for various SL configurations. Using the non-equilibrium Green's function (NEGF) formalism coupled self-consistently with the Poisson's equation,  we identify such an ideal structure and also demonstrate that it is almost immune to the deleterious effect of self-consistent charging and device variability. Analyzing various superlattice designs, we conclude that superlattices with a Gaussian distribution of the barrier thickness offers the best thermoelectric efficiency at maximum power. We also analyze our device designs with the conventional figure of merit approach to counter support the results so obtained. \\
\indent The paper is organized as follows.  In Sec. II, we describe the self-consistent simulation setup for numerical calculations and the details of the thermoelectric transport model. The simulation results are discussed in Sec. III, where we demonstrate in detail the process of selecting the best configuration of the SL structure keeping in mind the power and efficiency trade-off as well as the figure of merit considerations. Finally, in Sec. IV we conclude the article in favor of the optimal superlattice design for thermoelectric performance.
\section{Simulation Setup and Formulation}
Figure~\ref{Device}(a) shows a schematic of a typical thermoelectric generator set up, which consist of three sections: hot contact, cold contact, and the device region. Our focus is on the device region that effectively gives rise to an energy bandpass transmission lineshape \cite{Whitney2014} across it. We use GaAs-Al$_x$Ga$_{1-x}$As heterostructure system with mole fraction $x=0.1$ for the device region \cite{Agarwal2014}, because of its minimal variation in the lattice constants and effective masses, resulting in almost negligible variance in the band profile. The device regions we consider are sketched in Fig.~\ref{Device}(b), and are labeled as SLC-I, SLC-II, SLC-III and SLC-IV. The configuration SLC-I features a regular well-barrier structure in series along the transport $(\hat{z})$ direction. In Fig.~\ref{Device}(b), $w$ is the width of the well region, $b$ is the barrier thickness, and $\delta E_c$ is the barrier height. The configuration SLC-II is similar to SLC-I, but sandwiched between two barriers of half the thickness as that of the regular barriers, and serves as an anti-reflective (AR) region \cite{Pacher2001}. The configuration SLC-III features a Gaussian distribution of barrier heights \cite{Gomez1999}, with the middle barrier having maximum height $\delta E_{c_{max}}$. Similarly, SLC-IV features a Gaussian distribution of barrier thicknesses with $b_{max}$, the maximum thickness of the center barrier. The Gaussian variation is described as $exp[(k-i)^2/ \sigma^2]$, where $k$ is the barrier position, $i$ is the middle barrier considered in the configuration, and $\sigma$ is the variance of the Gaussian distribution.

The energy bandpass features can be attained via the structural configurations described above. For example, we can alter the transmissivity (area under the energy-transmission curve) by using the different configurations. We also note that the transmission in these SL structures is almost zero in the forbidden bands \cite{Tsu2010} thus potentially giving rise to the ``boxcar" like transmission feature \cite{Whitney2014}.

\begin{figure}
	\centering
	\includegraphics[width=0.5\textwidth]{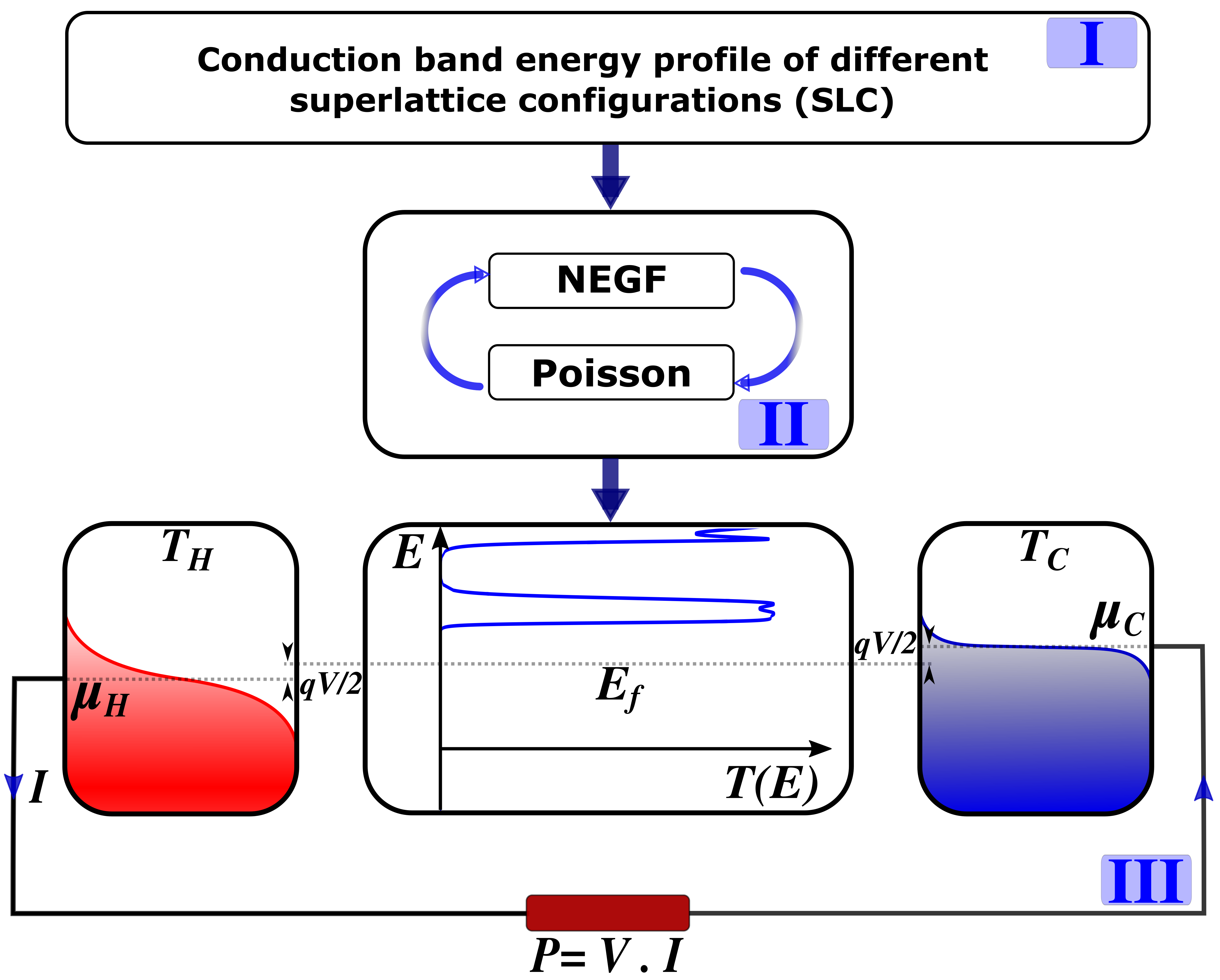}
	\caption{Simulation Flow chart (blockwise): (I) Conduction band energy offset $\delta E_c$, input from Fig.~\ref{Device}(b). (II) The barrier profile added with device Hamiltonian are solved self-consistently with NEGF-Poisson formulation, to generate the required transmission function $T(E)$. (III) is the voltage controlled TE generator set up. The transmission function is then fed into the Landauer equation, weighted by Fermi function difference ($f_H - f_C$), of hot and cold contacts with their Fermi energy $\mu_H$ and $\mu_C$ respectively, for the current $(I)$ calculation. Bias voltage $V$ is applied symmetrically to the average chemical potential $(E_f \geq 0)$ of the device at both the contacts. Power $P$ is given by the product of applied bias and current.}
	\label{NEGF_Box}
\end{figure}

The quantum transport description of the set up involves the device region described by the device  Hamiltonian $[H]$ and the coupling to the hot and cold contacts described via the self-energies $\Sigma_H$ and $\Sigma_C$ respectively \cite{QTDatta} as shown in Fig.~\ref{Device}(a). Figure~\ref{NEGF_Box} describes the flow of the simulation procedure. The description of the input conduction band profile is captured in Block-I. This carries all the information needed to characterize the different SLCs, using the matrix $[\delta E_c(z)]$, representing the potential profile of the device region. We introduce the nearest neighbor tight-binding model to form the effective-mass Hamiltonian matrix $[H]$, including the potential energy term $[\delta E_c(z)]$. We employ the coherent one band NEGF formulation \cite{QTDatta} to calculate transmission across the structures.

The NEGF equations start with the energy resolved retarded Green's function in its matrix representation $[G(E)]$, given by

\begin{equation}
[G(E)]=[E\mathbb{I}-H-U-\Sigma_H-\Sigma_C]^{-1},
\label{eqG}
\end{equation}

where $\mathbb{I}$ is the identity matrix, $\Sigma_{H(C)}$ is the self-energy matrix of hot (cold) contact. The combined effect of bias potential $(V)$ and electrostatic charging is encapsulated in the matrix $[U]$. It is obtained via a self-consistent calculation with the Poisson’s equation along the transport direction $(\hat{z})$ given by

\begin{equation}
\frac{d^2}{dz^2}(U(z)) = \frac{-q^2}{\epsilon_o \epsilon_r} n(z),
\label{eqPoisson}
\end{equation}

\begin{equation}
n(z) = \frac{1}{\Omega} \int\frac{G^n(E)}{2 \pi} dE,
\label{enz}
\end{equation}
where $n(z)$ is the electron density, $\Omega$ is the volume and $G^n$ is a diagonal element of the energy resolved electron correlation matrix $[G^n(E)]$ given by

\begin{equation}
[G^n(E)]=[G][\Gamma_{H} f_H + \Gamma_{C} f_C][G]^{\dagger},
\label{eqGn}
\end{equation}
where $\Gamma_{H(C)}$ represents the broadening matrices and $f_{H(C)}$ is the Fermi-Dirac distribution of the hot (cold) contact. We self consistently solve Eqs.~(\ref{eqG})-(\ref{eqGn}) to obtain the non-equilibrium transmission $T(V,E)$ given by

\begin{equation}
T(E)=Tr[\Gamma_{H} G \Gamma_{C} G^\dagger],
\label{eqTM}
\end{equation}
where $Tr$ denotes the Trace of the matrix. Using this we can now evaluate the charge current density $J$, by invoking the Landauer formula \cite{QTDatta}

\begin{multline}
J = \frac{2q}{h} \int dE\ T(V, E) \\ \times [F_{2D}(E-\mu_H) - F_{2D}(E-\mu_C)],
\label{eqJ}
\end{multline}

and the heat current densities

\begin{multline}
J_H^{Q1} = \frac{2}{h} \int dE\ T(V, E)\ (E-\mu_H) \\ \times [F_{2D}(E-\mu_H) - F_{2D}(E-\mu_C)],
\label{eqJH1}
\end{multline}

\begin{multline}
J_H^{Q2} = \frac{2}{h} \int dE\ T(V, E)\ (E-\mu_H) \\ \times [G_{2D}(E-\mu_H) - G_{2D}(E-\mu_C)],
\label{eqJH2}
\end{multline}
where $J_H^{Q1}$ and $J_H^{Q2}$ are the heat current densities in the longitudinal and transverse directions of transport respectively. The functions $F_{2D}$ and $G_{2D}$ are expressed as

\begin{equation}
F_{2D} = \frac{m_e^* k_B T} {2\pi\hbar^2} \log[1+exp(\frac {\mu-E} {k_BT})],
\label{eqF2D}
\end{equation}

\begin{equation}
G_{2D} = \frac{m_e^* k_B T} {2\pi\hbar^2} \int dE \frac {E_{\perp}}{1+exp(\frac {E+E_{\perp}-\mu} {k_BT})},
\label{eqG2D}
\end{equation}
where $q$ is the electronic charge, $h$ is the Planck's constant in $eV.S$, $k_B$ is the Boltzmann's constant. These integrals are due to transverse mode summations\cite{Sothmann2013}. The total heat current density is then given by $J_H^{Q} = J_H^{Q1} + J_H^{Q2}$.

The calculated charge current density $(J)$ is used to obtain the output power density $P=-JV$, from which the efficiency can be obtained as $\eta = P/J_H^Q$. The conversion efficiency is calculated with respect to the Carnot efficiency (i.e. $\eta/\eta_C$). We assume a symmetric electrostatic coupling to the contacts due to which a bias voltage $V$ results in a chemical potential shift of $\mp qV/2$ at the contacts as seen in Block-II of Fig.~\ref{NEGF_Box}. At a particular voltage, referred to as the open circuit voltage $V_{oc}$, the charge current completely opposes that which is set up by the thermal gradient, also known as the Seebeck voltage. In the region $[0, V_{oc}]$, the set up works as a thermoelectric generator.

\section{Results and Discussion}
In the following, we describe simulation results starting from the calculation of transmission function $T(E)$, to various TE performance parameters of SLC-(I to IV). 

\subsection{Performance Analysis of SLC-I and SLC-II}

\begin{figure}
	\subfigure[]{\includegraphics[height=0.18\textwidth,width=0.225\textwidth]{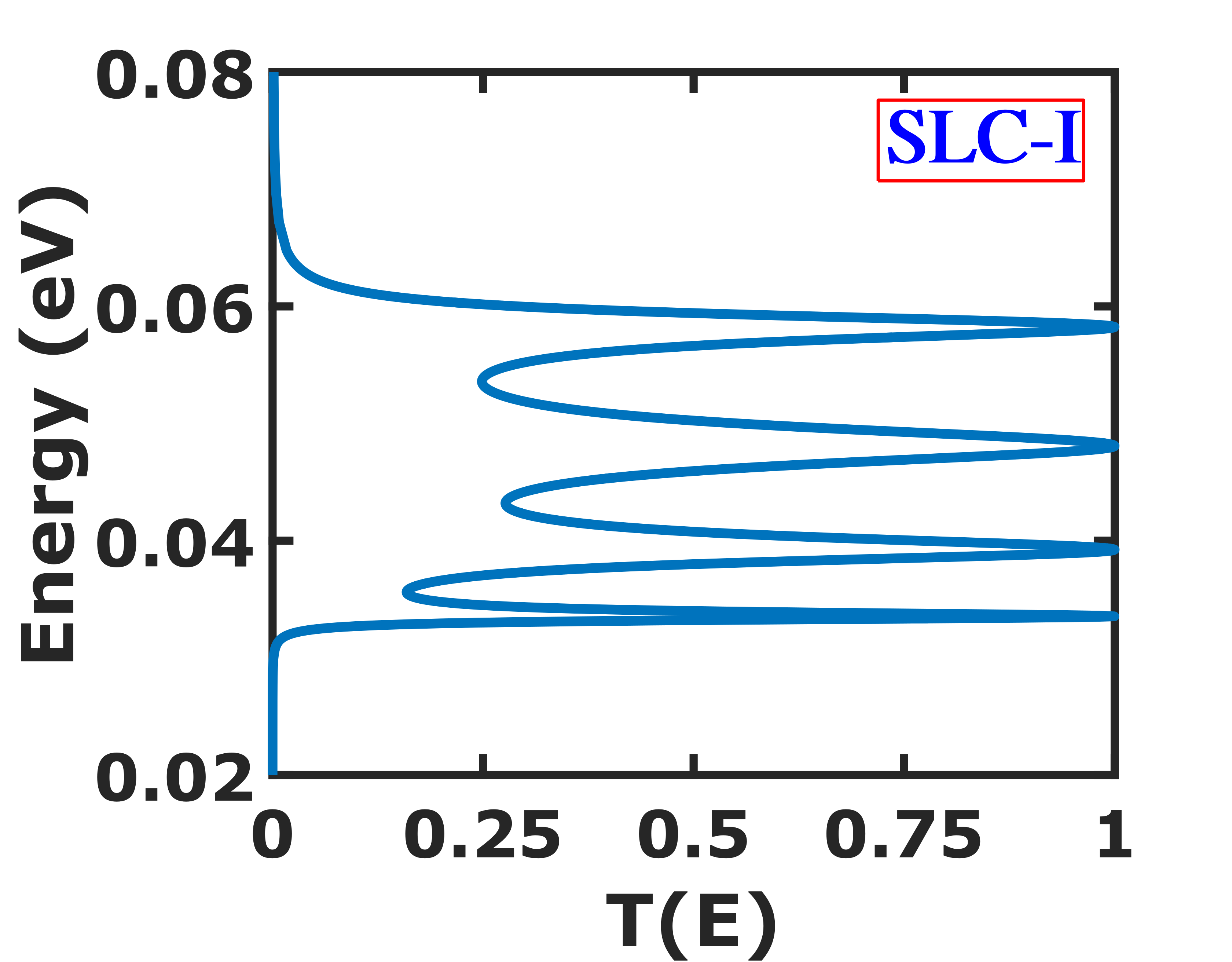}\label{5B_NSL_TM_E}}
	\quad
	\subfigure[]{\includegraphics[height=0.18\textwidth,width=0.225\textwidth]{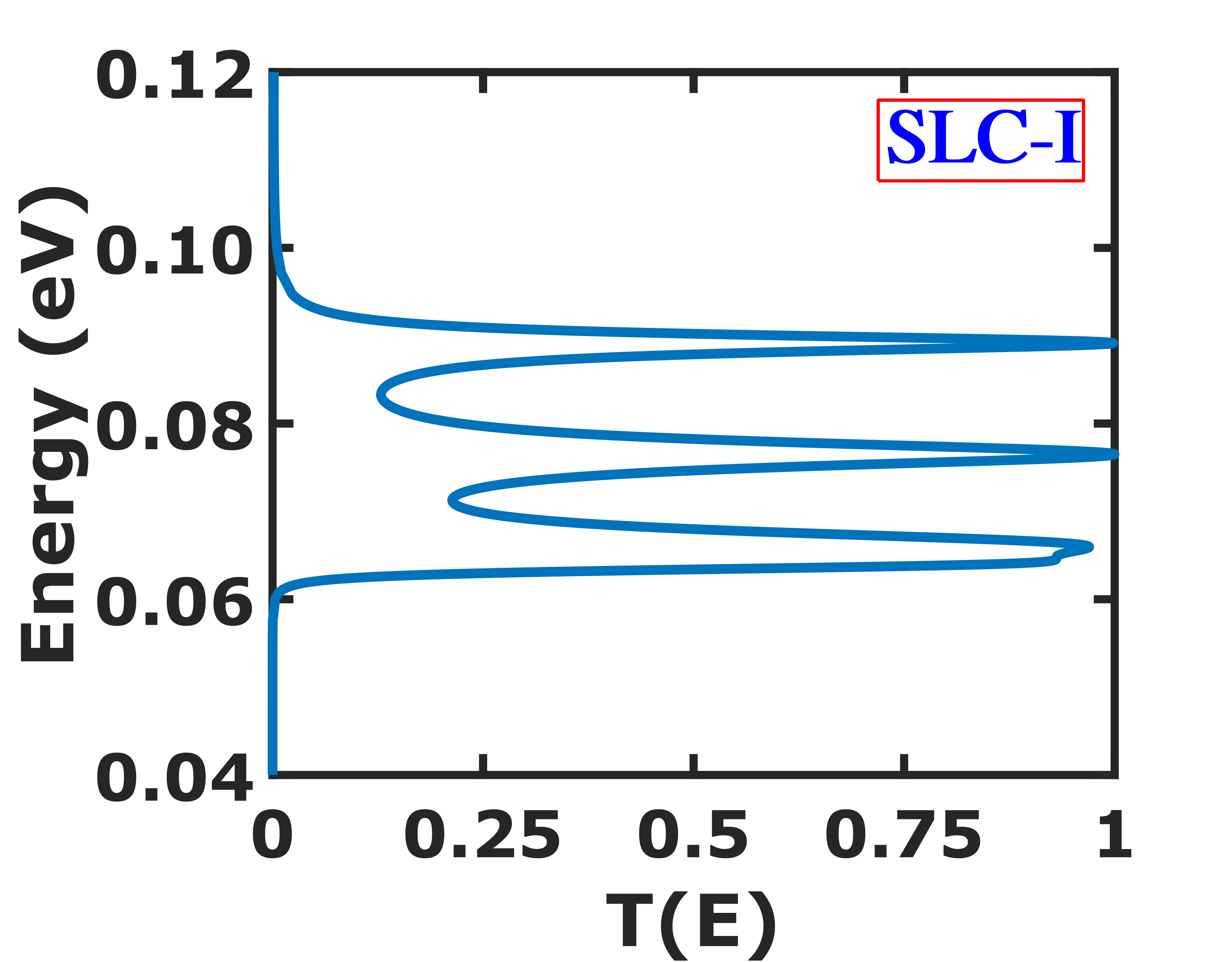}\label{5B_NSL_TM_E_Poisson}}
	\quad
	\subfigure[]{\includegraphics[height=0.18\textwidth,width=0.225\textwidth]{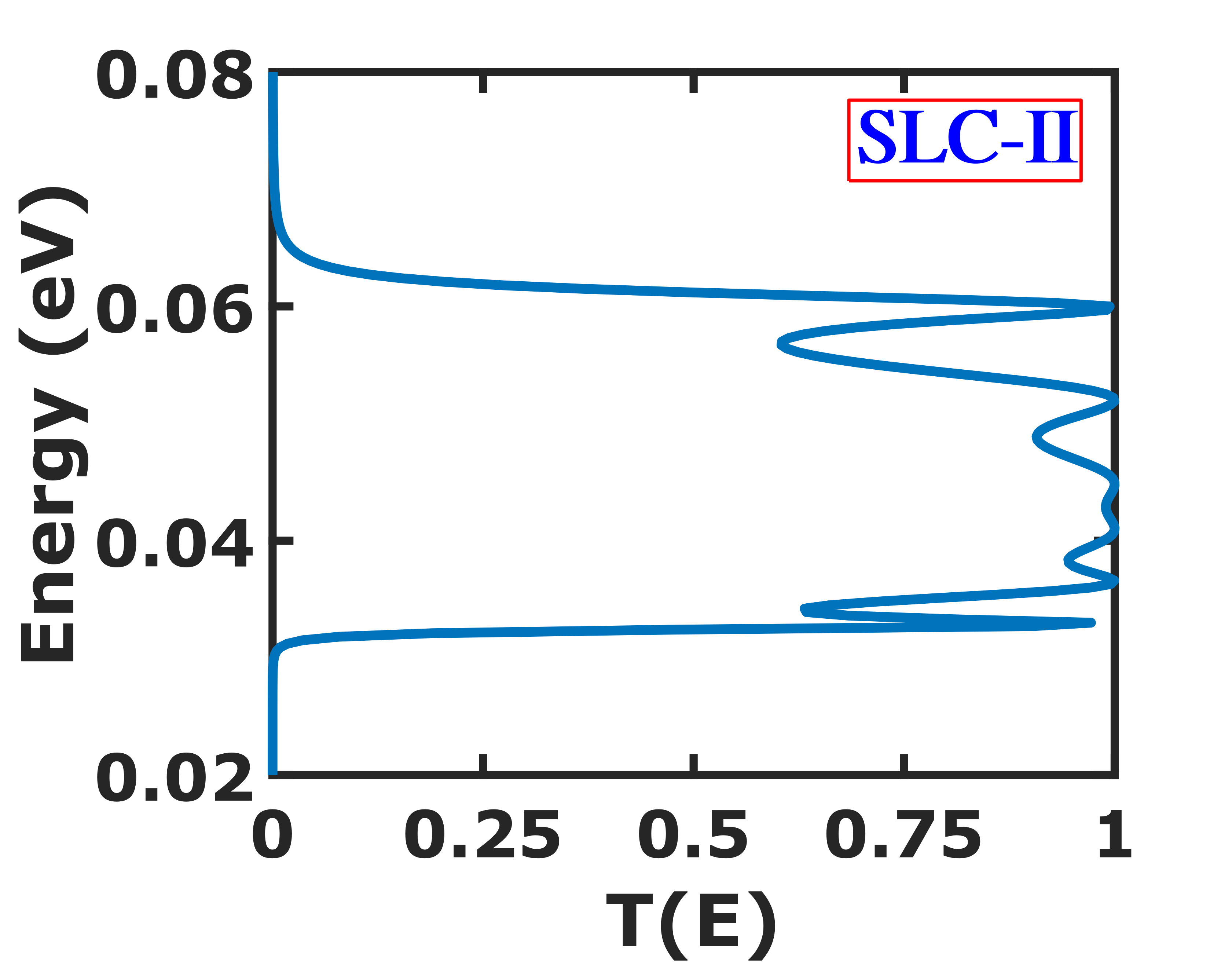}\label{5B_ARC_TM_E}}
	\quad
	\subfigure[]{\includegraphics[height=0.18\textwidth,width=0.225\textwidth]{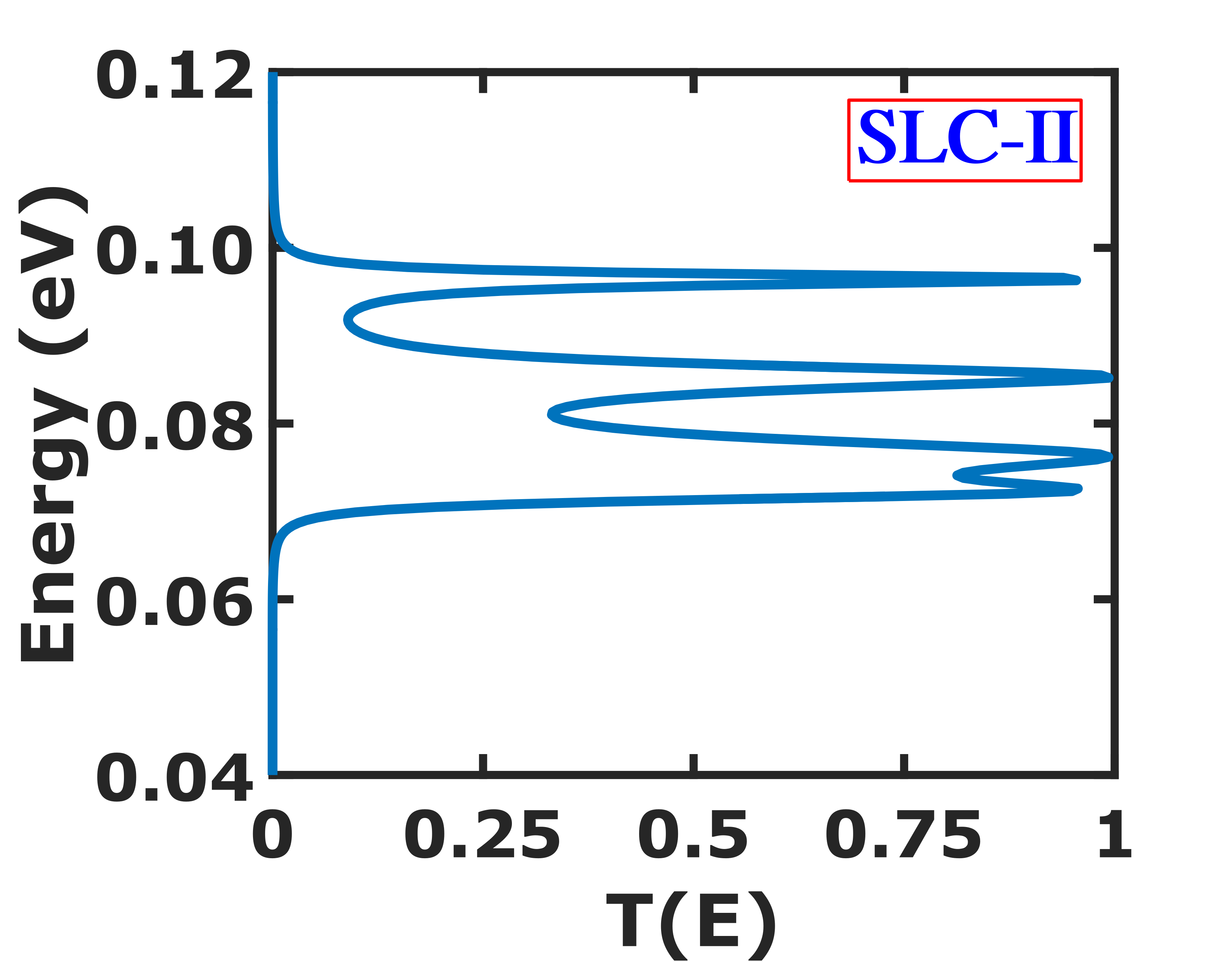}\label{5B_ARC_TM_E_Poisson}}
	
	\caption{Transmission coefficient as a function of energy $T(E)$ of the configuration type (a) SLC-I and (c) SLC-II, at zero applied bias for the 5-barriers without the inclusion of the Poisson's equation. The transmission, $T(E)$ of (b) SLC-I and (d) SLC-II, evaluated with self-consistent Poisson keeping $V=0$ and $E_f=0~k_BT$. Plots are zoomed in only for the first miniband.}
	\label{5B_TM_E}
\end{figure}

\begin{figure}
	\centering
	\subfigure[]{\includegraphics[height=0.18\textwidth,width=0.225\textwidth]{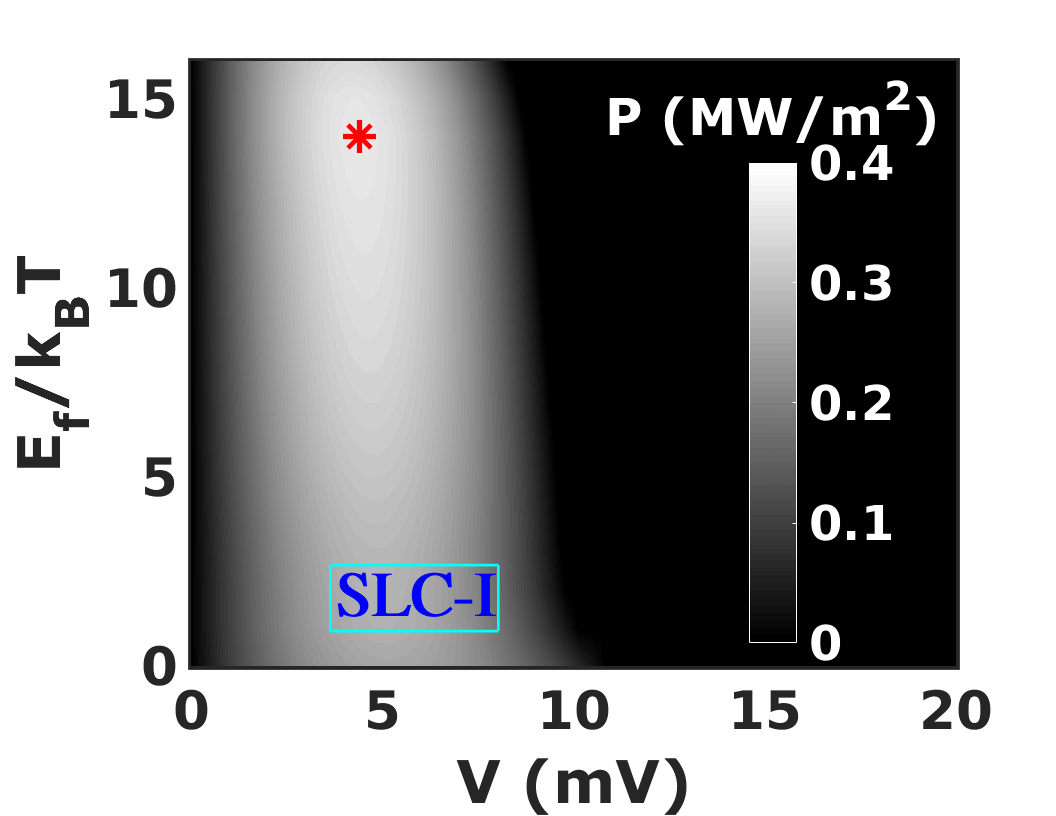}\label{5B_NSL_P_V_kT}}
	\quad
	\subfigure[]{\includegraphics[height=0.18\textwidth,width=0.225\textwidth]{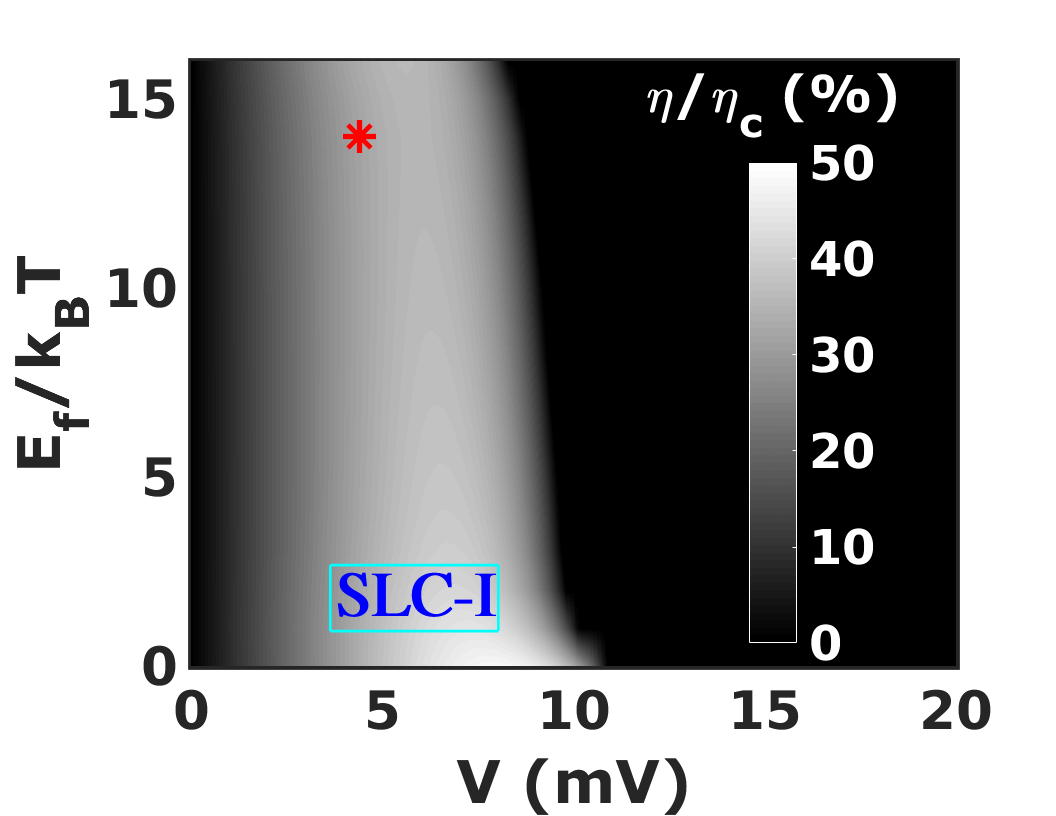}\label{5B_NSL_Eff_V_kT}}
	\quad
	\subfigure[]{\includegraphics[height=0.18\textwidth,width=0.225\textwidth]{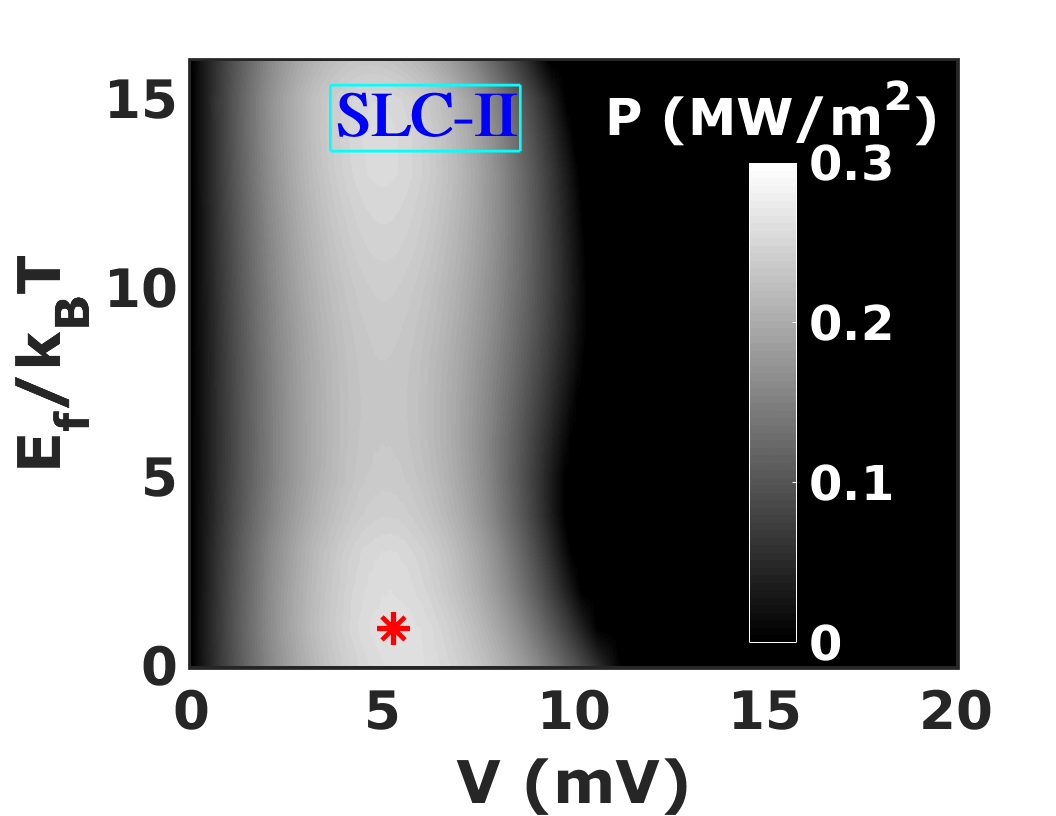}\label{5B_ARC_P_V_kT}}
	\quad
	\subfigure[]{\includegraphics[height=0.18\textwidth,width=0.225\textwidth]{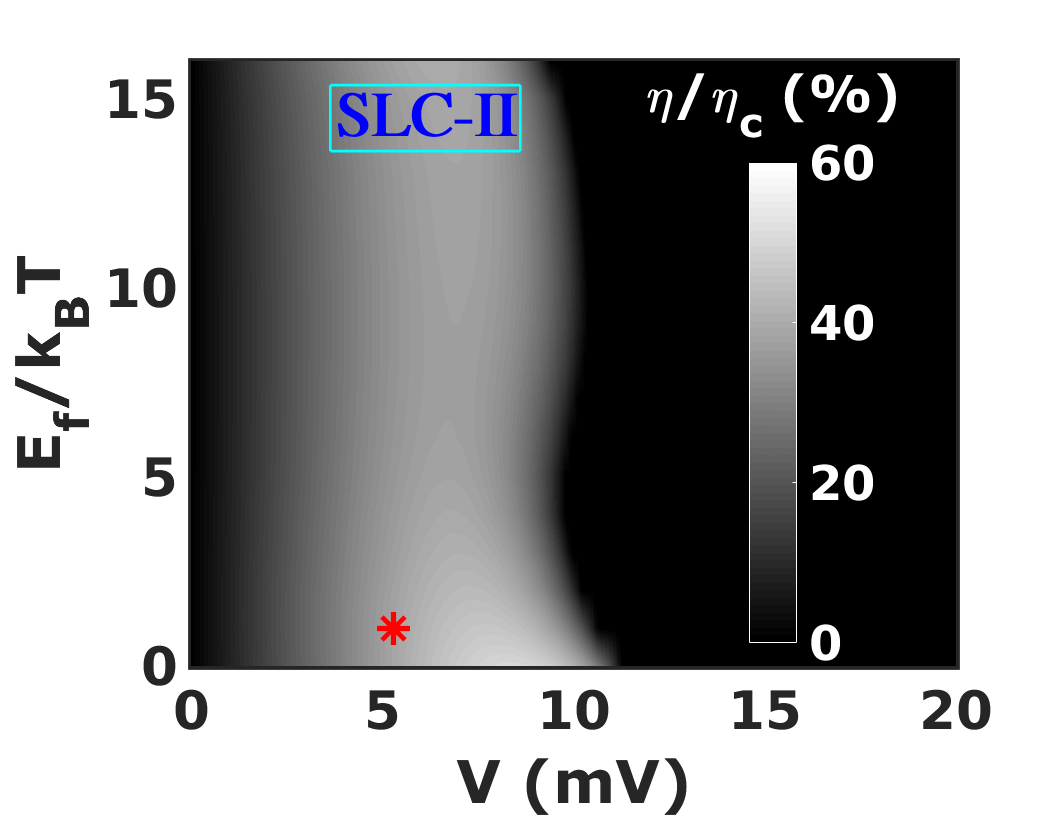}\label{5B_ARC_Eff_V_kT}}
	
	\caption{Bar plot: Power density (P in $MW/m^2$) and efficiency $\eta / \eta_c$ plotted on a gray scale as a function of applied bias $V$ and average electrochemical potential $E_f$ for SLC-I (a \& b) and SLC-II (c \& d). Maximum output power and corresponding normalized efficiency are marked red $(\ast)$.} 
	\label{5B_P_Eff}
\end{figure}

\emph{Miniband Formation}: In the structures considered with 5 regular barriers, we set the well width as 6 $nm$ and the barrier thickness as 4 $nm$ with a height of 0.1 $eV$. The transmission calculations are performed with $m^{\star}_e=0.07m_o$, where $m_o$ is the free electron mass.  The standard solution of the Schr\"{o}dinger equation leads to the formation of minibands. 

Focusing on SLC-I in Fig.~\ref{5B_NSL_TM_E}, we note that the number of peaks that originate in the lowest miniband depends on the number of wells present in the structure. However, the inclusion of Poisson equation, as shown in Fig.~\ref{5B_NSL_TM_E_Poisson}, lift up the band center in energy and distorts the peaks. The peaked feature of $T(E)$ may in general not desired for the thermoelectric power generation as pointed out by Whitney\cite{Whitney2014}. We can obtain a closed to ``boxcar" transmission feature using SLC-II, as seen in Fig.~\ref{5B_ARC_TM_E}, which enables an AR region \cite{Pacher2001, Morozov2002}. However, interestingly as seen in Fig.~\ref{5B_ARC_TM_E_Poisson}, the boxcar feature and hence the utility of the AR region is completely destroyed when the Poisson solution is applied.

The above analysis shows that the AR structures proposed by Karbaschi\cite{Karbaschi2016}, are not of much utility in a realistic situation when we take charging effect into account. We also observe that although the transmitivity can increase in SLC-II due to the AR region, it shows a degradation when Poisson charging is taken into account.

We now turn our attention to the thermoelectric performance evaluated via the power-efficiency analysis for the configurations, SLC-I and SLC-II. In Fig.~\ref{5B_P_Eff}, we plot the calculated power density and efficiency (normalized with $\eta_C$) as a function of the applied bias $V$ for different values of the electrochemical potential or Fermi level $E_f$, with Poisson charging taken into account. To achieve a thermoelectric effect, the Fermi level $E_f$, should be kept below the bottom of the first miniband. Varying $E_f$ from its zero value up to it crossing with the lowest energy of the miniband. We find that there exists a point at which the output power is maximized. The shaded area in Fig.~\ref{5B_P_Eff} shows the working region of the thermoelectric power generator.

Figure~\ref{5B_NSL_P_V_kT} and \ref{5B_NSL_Eff_V_kT} depict the power and efficiency plot respectively for SLC-I, where the red asterisk $(\ast)$  denotes the operating point of maximum achievable power. The corresponding efficiency at maximum power is also similarly marked in Fig.~\ref{5B_NSL_Eff_V_kT}. The maximum output power density of 0.36 $MW/m^2$ is achieved at $E_f$=14$k_B T$ for SLC-I, with a corresponding efficiency of 32 \% of the Carnot value. Likewise, the power and efficiency plots in Fig.~\ref{5B_ARC_P_V_kT} and \ref{5B_ARC_Eff_V_kT} for SLC-II show that the maximum power is 0.26 $MW/m^2$ at $E_f$=$k_B T$. Therefore we can conclude from the above analysis that SLC-II, namely the AR superlattice performance, in fact, degrades when the charging effect is taken into consideration.

\subsection{Performance Analysis of SLC-III and SLC-IV}

\begin{figure}
	\subfigure[]{\includegraphics[height=0.18\textwidth,width=0.225\textwidth]{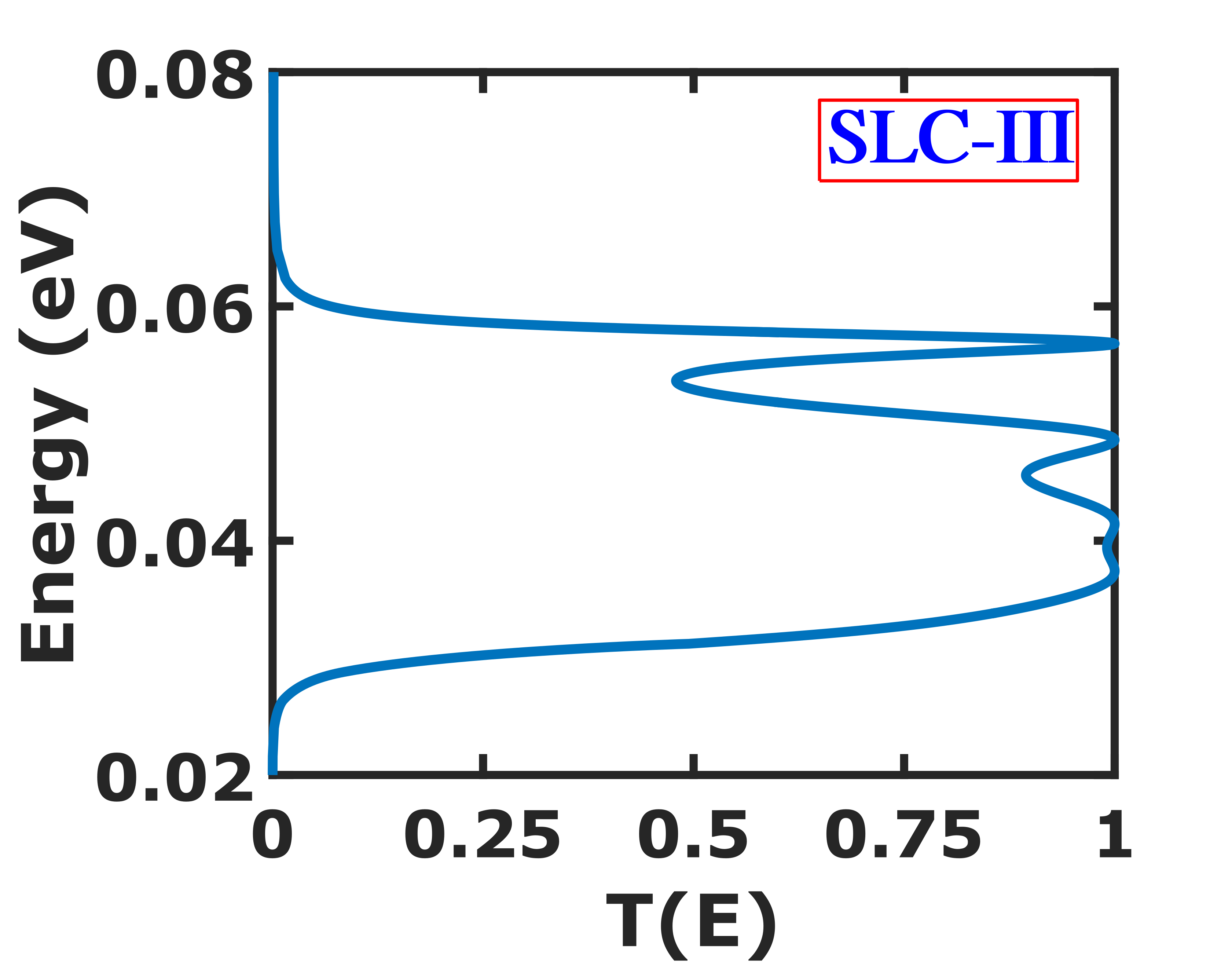}\label{11B_GEc_TM_E}}
	\quad
	\subfigure[]{\includegraphics[height=0.18\textwidth,width=0.225\textwidth]{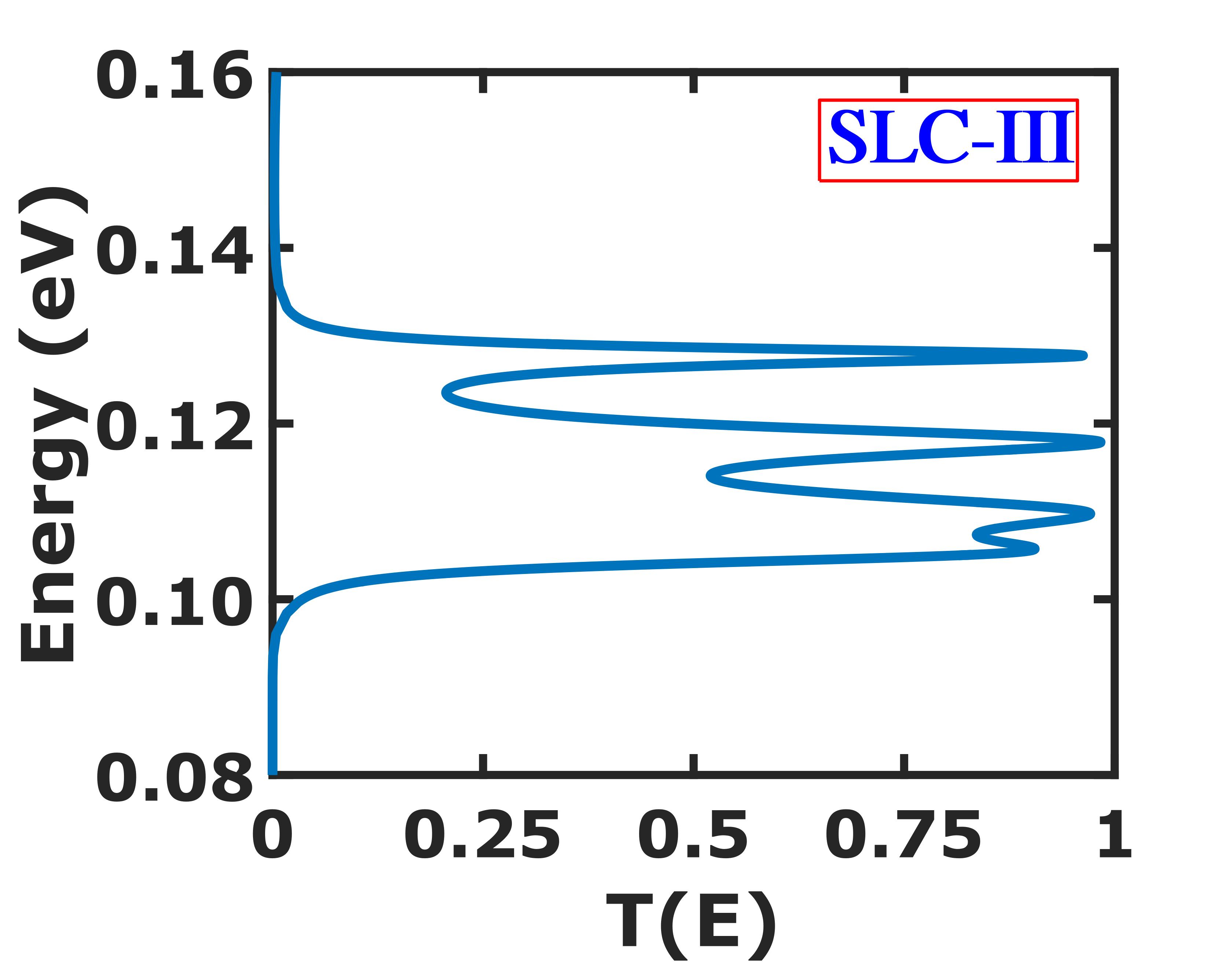}\label{11B_GEc_TM_E_Poisson}}
	\quad
	\subfigure[]{\includegraphics[height=0.18\textwidth,width=0.225\textwidth]{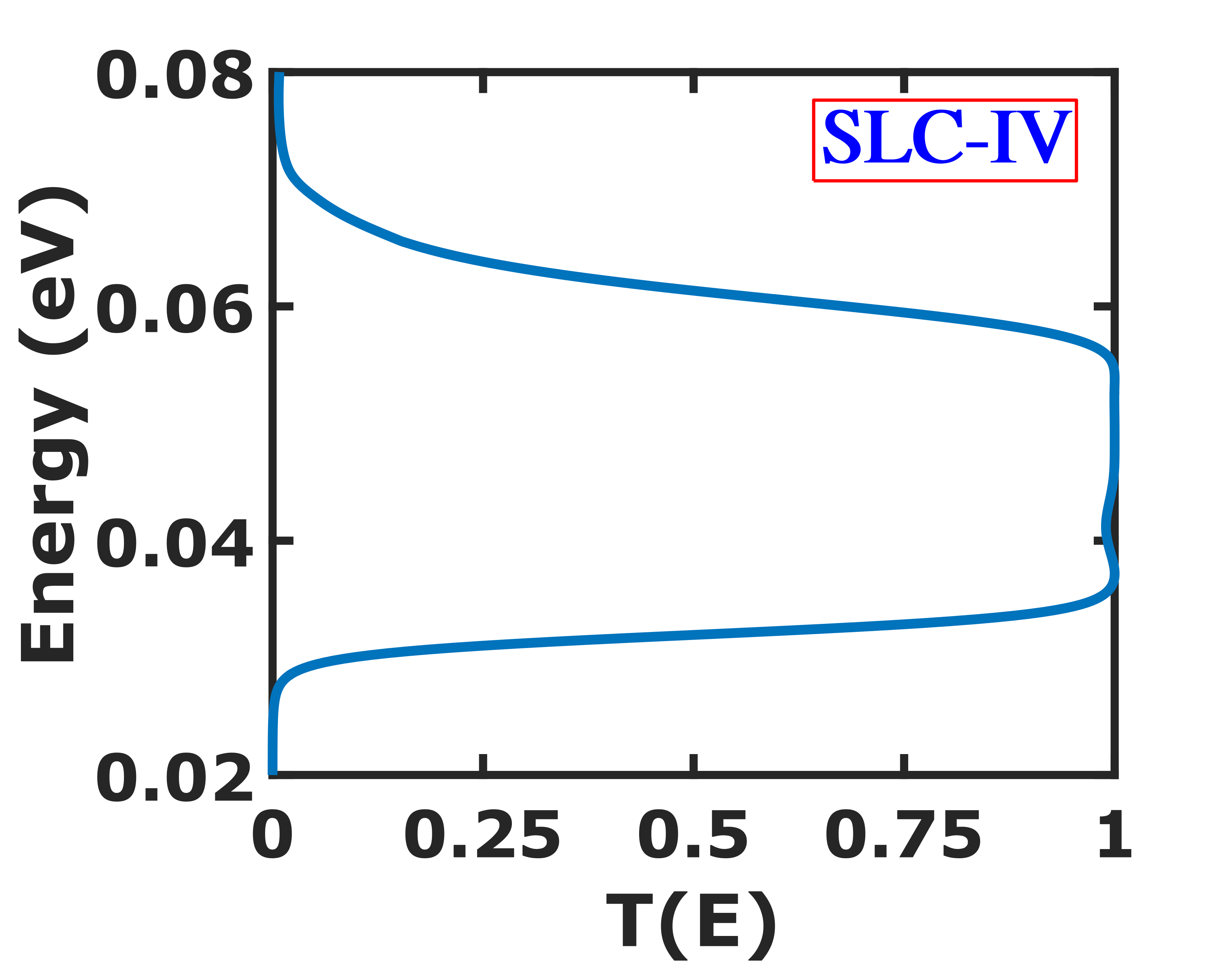}\label{11B_GW_TM_E}}
	\quad
	\subfigure[]{\includegraphics[height=0.18\textwidth,width=0.225\textwidth]{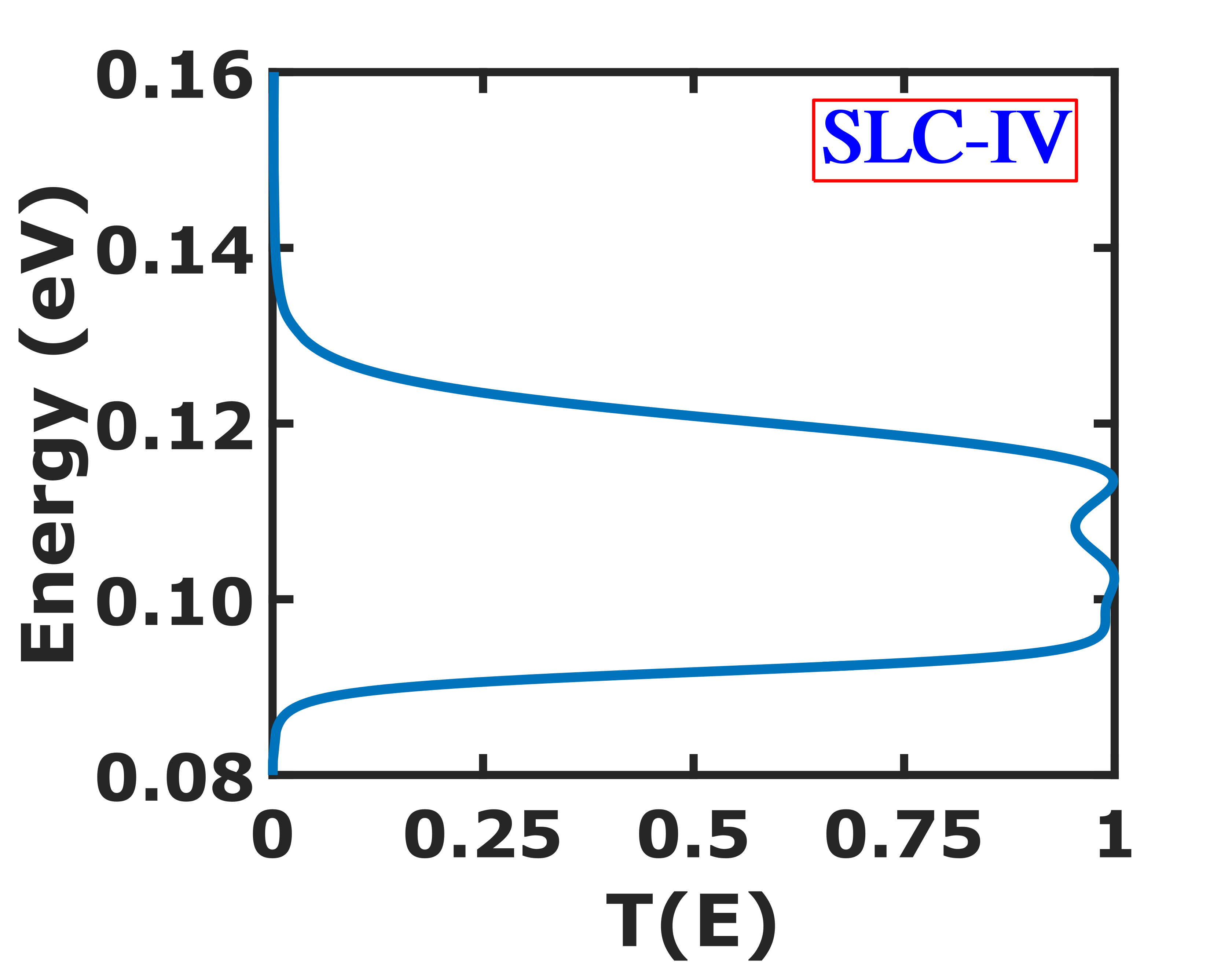}\label{11B_GW_TM_E_Poisson}}
	
	\caption{Transmission coefficient as a function of energy $T(E)$ of the configuration type (a) SLC-III and (c) SLC-IV, at zero applied bias for the 11-barrier structure without the inclusion of Poisson charging. $T(E)$ of (b) SLC-III and (d) SLC-IV, evaluated with self consistent Poisson keeping, $V=0$ and $E_f=0~k_BT$. Plots are zoomed in only for the first miniband.}
	\label{11B_TM_E}
\end{figure}

\begin{figure}
	\centering
	\subfigure[]{\includegraphics[height=0.18\textwidth,width=0.225\textwidth]{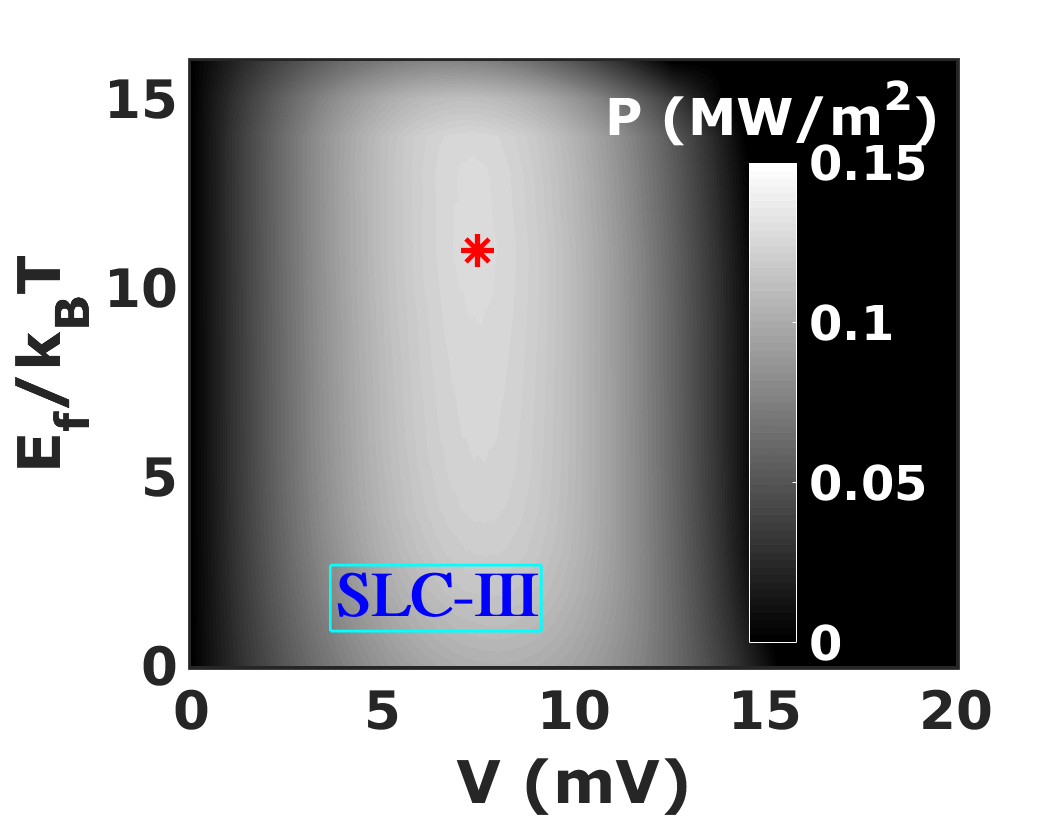}\label{11B_GEc_P_V_kT}}
	\quad
	\subfigure[]{\includegraphics[height=0.18\textwidth,width=0.225\textwidth]{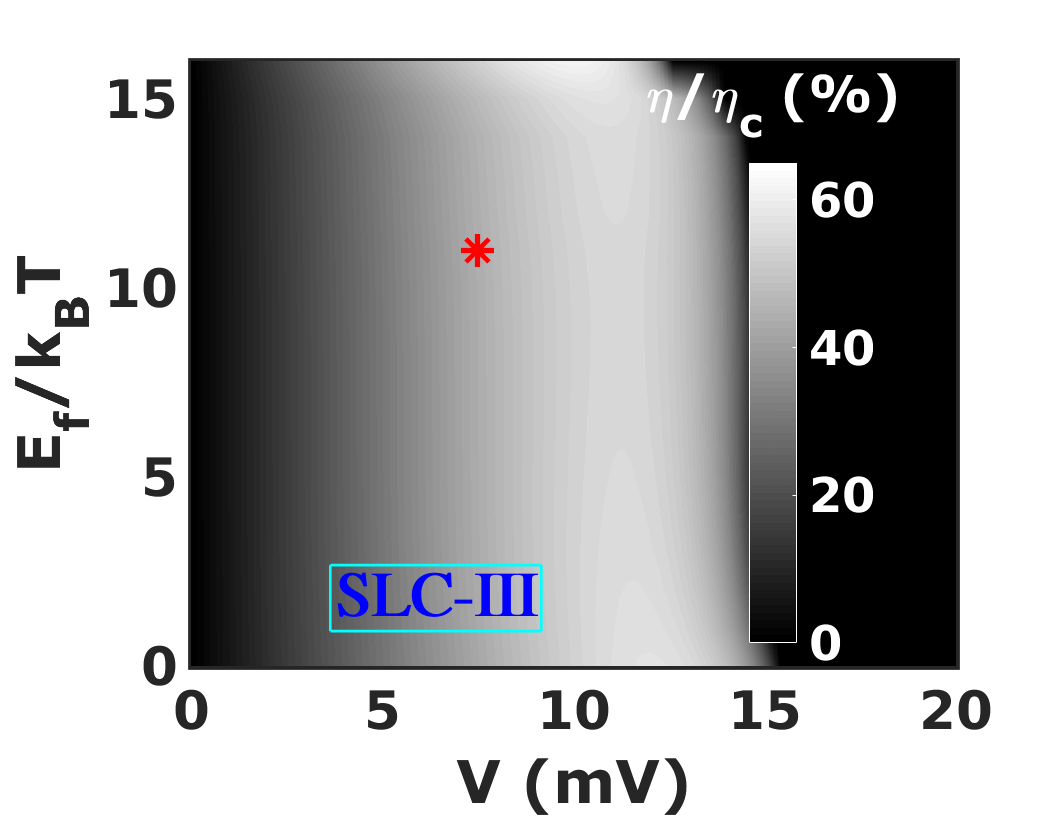}\label{11B_GEc_Eff_V_kT}}
	\quad
	\subfigure[]{\includegraphics[height=0.18\textwidth,width=0.225\textwidth]{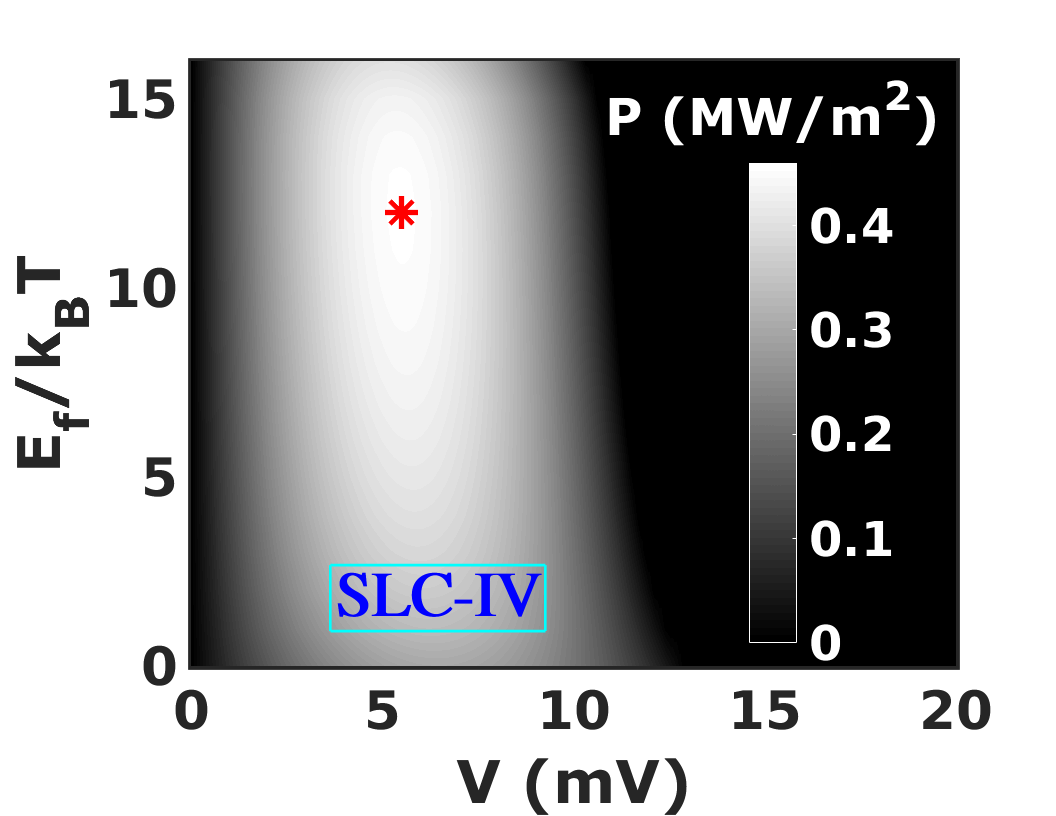}\label{11B_GW_P_V_kT}}
	\quad
	\subfigure[]{\includegraphics[height=0.18\textwidth,width=0.225\textwidth]{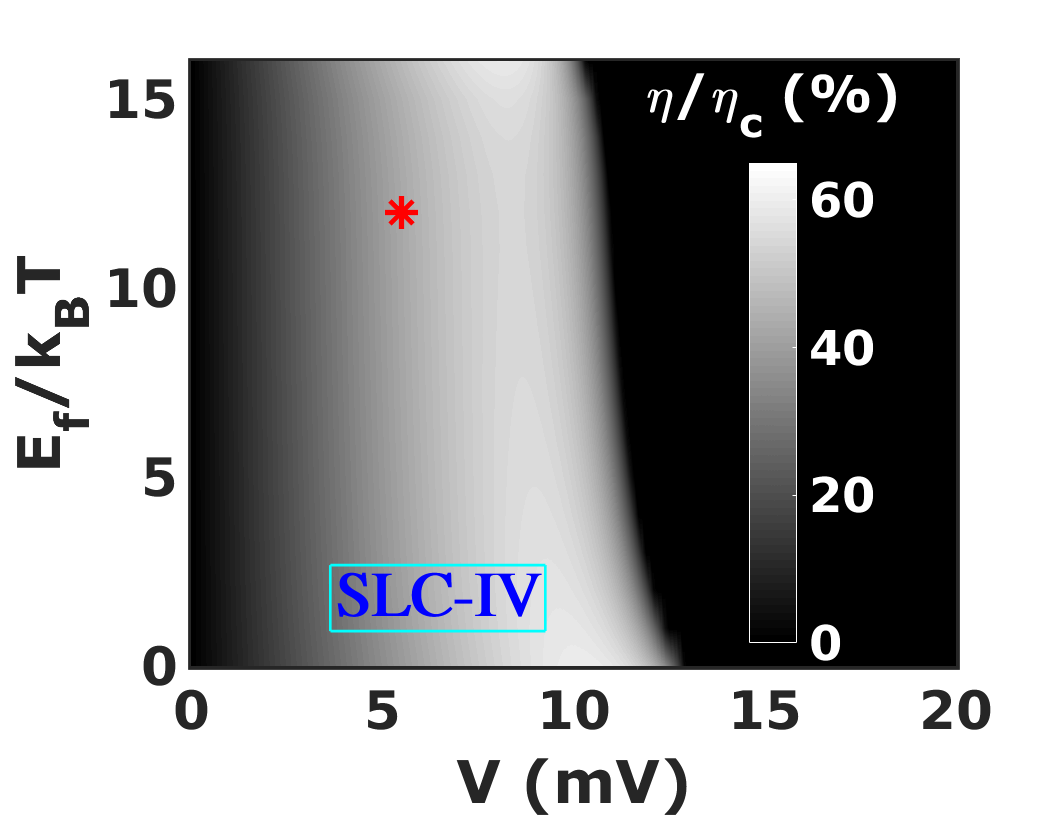}\label{11B_GW_Eff_V_kT}}
	
	\caption{Bar plot: Power density (P in $MW/m^2$) and efficiency $\eta / \eta_C$ plotted on a gray scale as a function of applied bias $V$ and average electrochemical potential $E_f$ for SLC-III (a \& b) and SLC-IV (c \& d). Maximum output power and corresponding normalized efficiency are marked red $(\ast)$.} 
	\label{11B_P_Eff}
\end{figure}

We now consider the Gaussian distributed superlattices, i.e. SLC-III and SLC-IV defined in Fig.~\ref{Device}, in the device region. The SLC-III is constructed using 11 barriers with the height of the $k^{th}$ barrier is given as $\delta E_c^k=\delta E_{c_{max}}exp[(k-6)^2/0.125]$, where $\delta E_{c_{max}}$ is the maximum height of the middle barrier taken as 0.1 $eV$. Likewise, 11 barriers are used in constructing SLC-IV, where we now vary the thickness of the barrier instead of its height. The thickness of the $k^{th}$ barrier is given by $b^k=b_{max}exp[(k-6)^2/2]$, where $b_{max}= 4~nm$, is the maximum thickness assigned to the middle barrier.

In Fig.~\ref{11B_GEc_TM_E}, we plot the transmission of the first miniband for SLC-III. Once again we note from Fig.~\ref{11B_GEc_TM_E_Poisson} that the transmission peaks get distorted with the inclusion of Poisson charging. Similarly Figs.~\ref{11B_GW_TM_E} and \ref{11B_GW_TM_E_Poisson} show the transmission plots for SLC-IV without and with Poisson charging respectively.

It turns out, upon comparing the $T(E)$ plots of SLC-III (Figs.~\ref{11B_GEc_TM_E} \& \ref{11B_GEc_TM_E_Poisson}) and SLC-IV (Figs.~\ref{11B_GW_TM_E} \& \ref{11B_GW_TM_E_Poisson}), that the latter is more immune to the charging effects, with transmission coefficient in SLC-IV case, being close to unity in the entire miniband, we expect a larger transmissivity. Thus among all superlattice configurations considered, we note that SLC-IV features the best ``boxcar" type transmission feature even with the inclusion of realistic charge effects.

We now evaluate the output power and efficiency of SLC-III and SLC-IV to analyze the thermoelectric performance. The power density and efficiency plots are shown in Figs.~\ref{11B_GEc_P_V_kT} and \ref{11B_GEc_Eff_V_kT} for the SLC-III case and in Figs.~\ref{11B_GW_P_V_kT} and \ref{11B_GW_Eff_V_kT} for the SLC-IV case. We note in the SLC-III case, that surprisingly the maximum power density is very low $(\approx 0.13~MW/m^2)$ at $E_f= 11~k_BT$. The efficiency at maximum power is $44 \%$ of the Carnot value. However, in the case of SLC-IV, we find that the maximum power is $0.46~MW/m^2$ at $E_f=12~k_BT$ at an efficiency of $43~\%$ of the Carnot value. We thus note that SLC-IV outperforms all the superlattice device structures discussed so far. We summarize the performance of all devices in Table~\ref{table:1}. In all SLCs we find the maximum efficiency $(\eta_{max})$ at $E_f=0~k_BT$.

\begin{table}
\centering
\caption{TE performance analysis of SLCs devices.}
	\begin{tabular}{c|c|c|c} 
		\hline
		SL Configuration & $P_{max}(MW/m^2)$ & $\eta_{P_{max}}(\%)$ & $\eta_{max}(\%)$ \\ 
		\hline\hline
		SLC-I & 0.36 & 31.78 & 48.14 \\ 
		\hline
		SLC-II & 0.26 & 38.46 & 50.16 \\
		\hline
		SLC-III & 0.13 & 44 & 61.70 \\
		\hline
		SLC-IV & 0.46 & 43 & 59.68 \\
		\hline
	\end{tabular}

\label{table:1}
\end{table}

\subsection{Power-efficiency Trade-off}
\begin{figure}[h]
	\centering
	{\includegraphics[height=0.32\textwidth,width=0.45\textwidth]{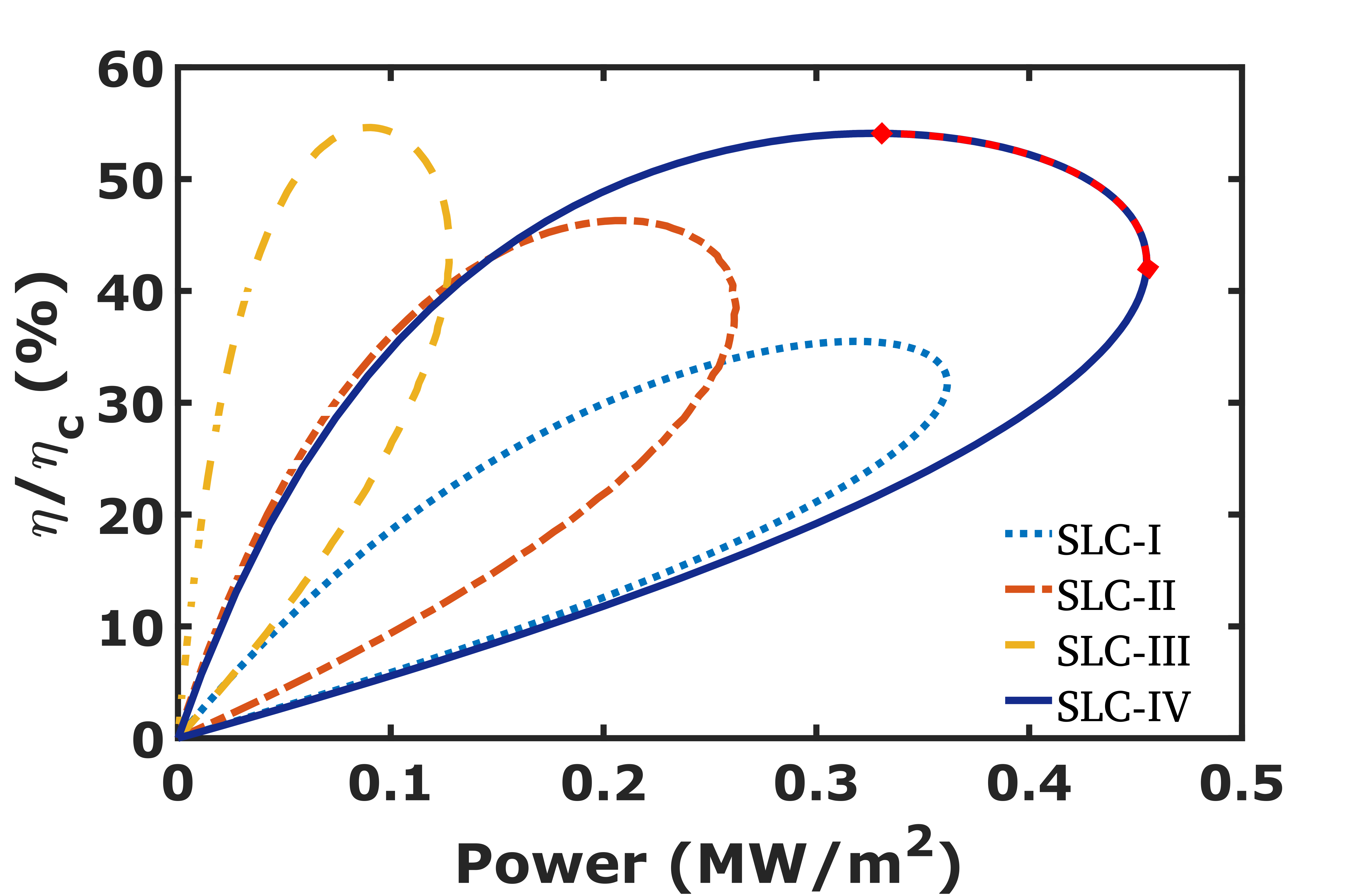}}
	
	\caption{Color loop: Power-efficiency trade-off at a particular value of $E_f$, where the corresponding power is maximum in all the four configurations. Efficiency is plotted in \% normalized to the Carnot efficiency $\eta_c$.}	
	\label{P_Eff_Tradeoff}
\end{figure}

Generally, the maximum conversion efficiency is achieved at the cost of a smaller output power and vice-versa\cite{Muralidharan2012}. Therefore, an operating point for an ideal TE performance should represent the optimization of the power-efficiency trade-off. In Fig.~\ref{P_Eff_Tradeoff}, we show the power-efficiency trade-off for all four device configurations at a Fermi energy where $P_{max}$ was achieved in each case. We note that SLC-I and SLC-II show less optimized power and efficiency despite using less number of barriers. Increasing the number of barriers in either configuration results in a drastic reduction in transmissivity and hence the power. The maximum efficiency is attained for the SLC-III device, $61 \%$ of the Carnot efficiency at a very low output power. The best results are obtained in the range of $0.32 - 0.46~ MW/m^2$ with efficiencies between $54 \% - 43 \%$ for the SLC-IV structure (solid blue loop with red dotted points in Fig.~\ref{P_Eff_Tradeoff}).

\begin{figure}
	\centering
	\subfigure[]{\includegraphics[height=0.18\textwidth,width=0.225\textwidth]{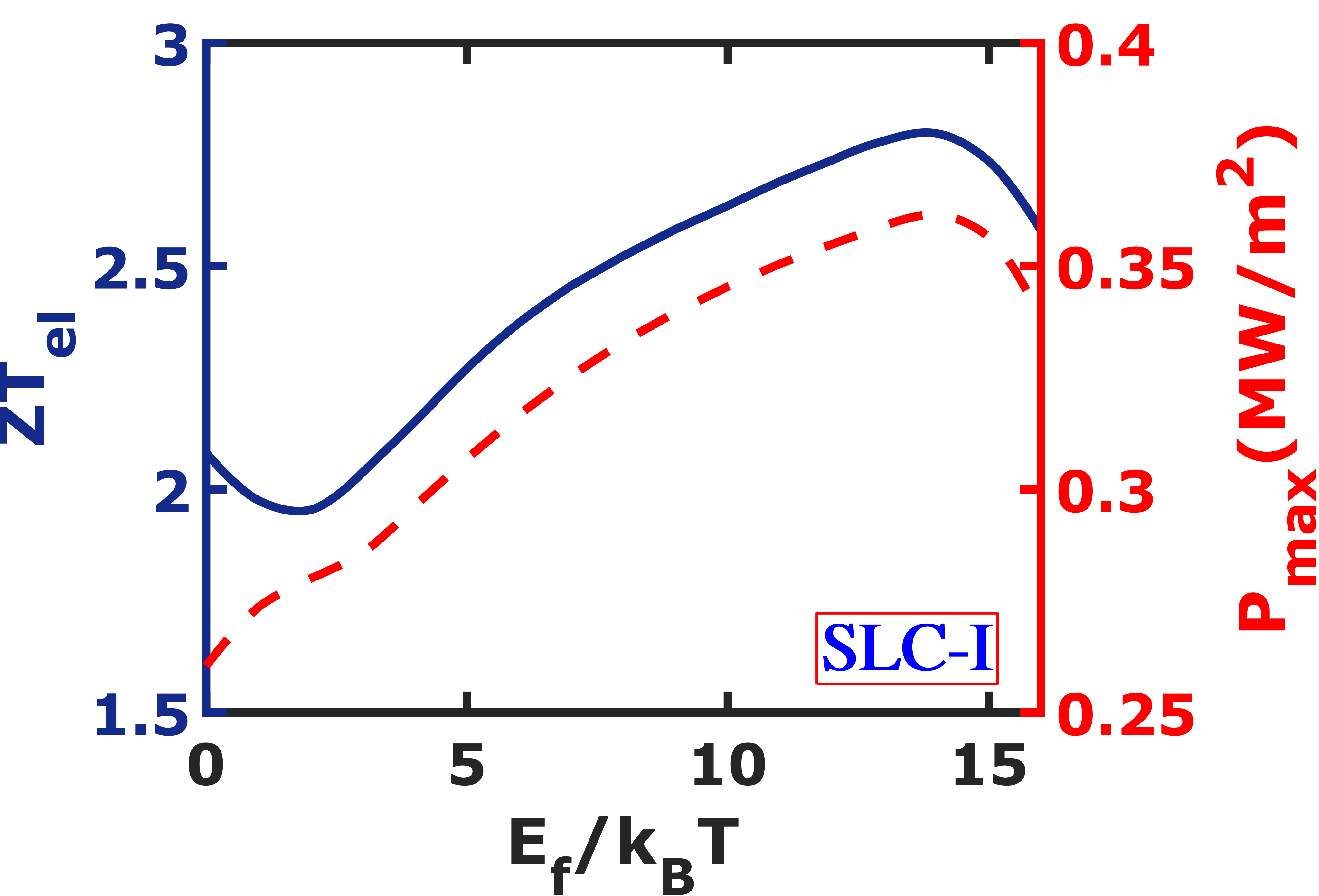}\label{5B_NSL_ZT_Pmax_kT}}
	\quad
	\subfigure[]{\includegraphics[height=0.18\textwidth,width=0.225\textwidth]{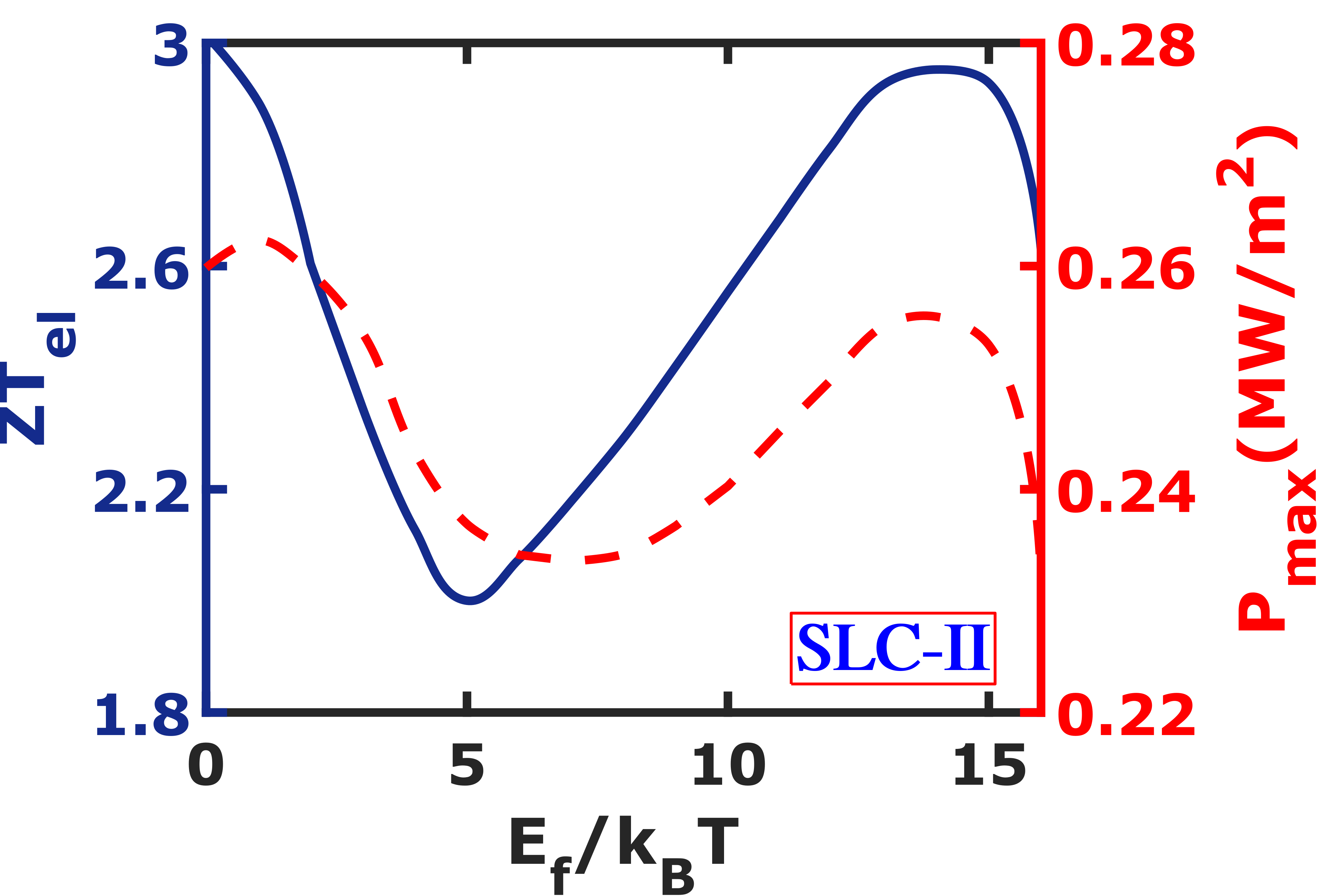}\label{5B_ARC_ZT_Pmax_kT}}
	\quad
	\subfigure[]{\includegraphics[height=0.18\textwidth,width=0.225\textwidth]{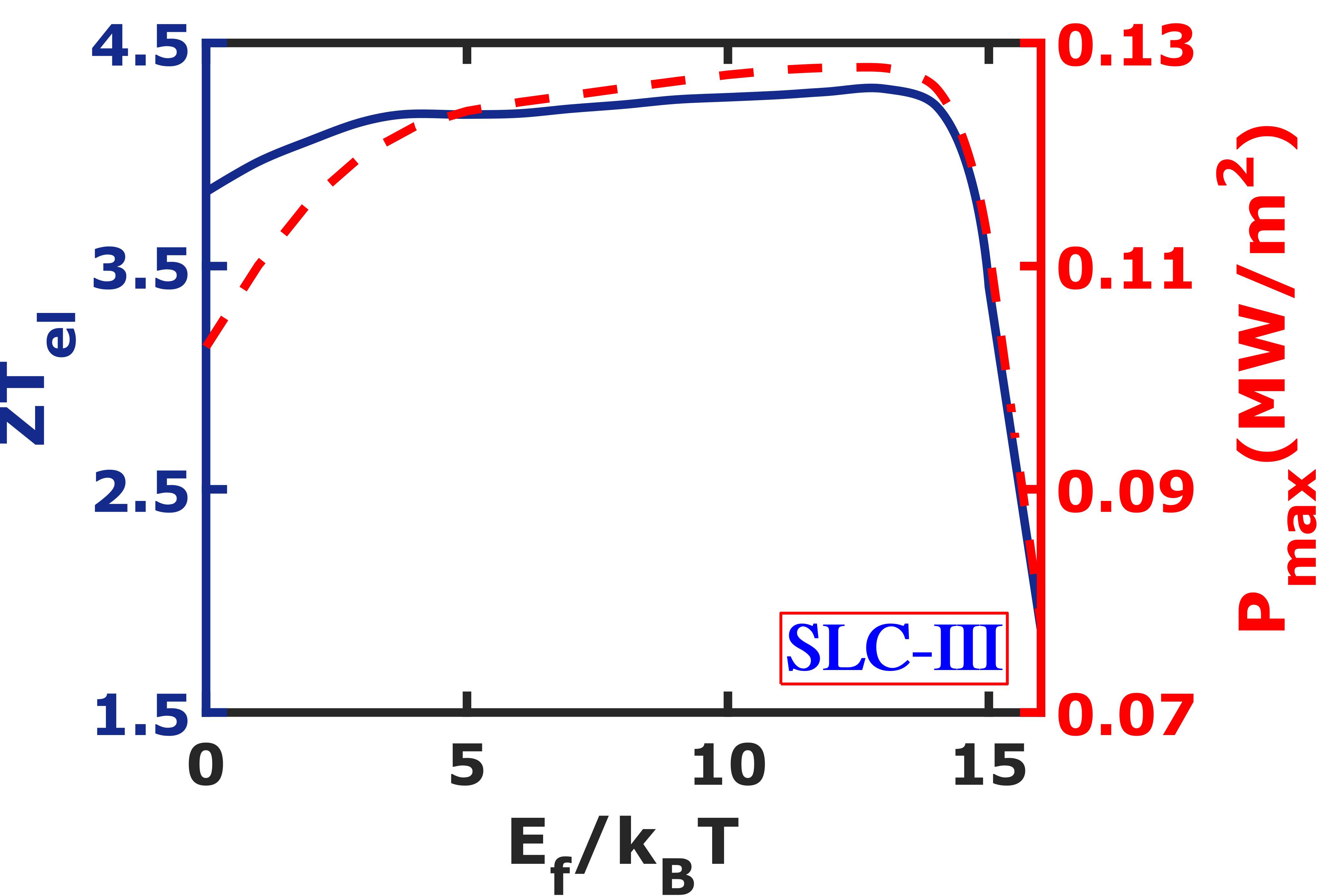}\label{11B_GEc_ZT_Pmax_kT}}
	\quad
	\subfigure[]{\includegraphics[height=0.18\textwidth,width=0.225\textwidth]{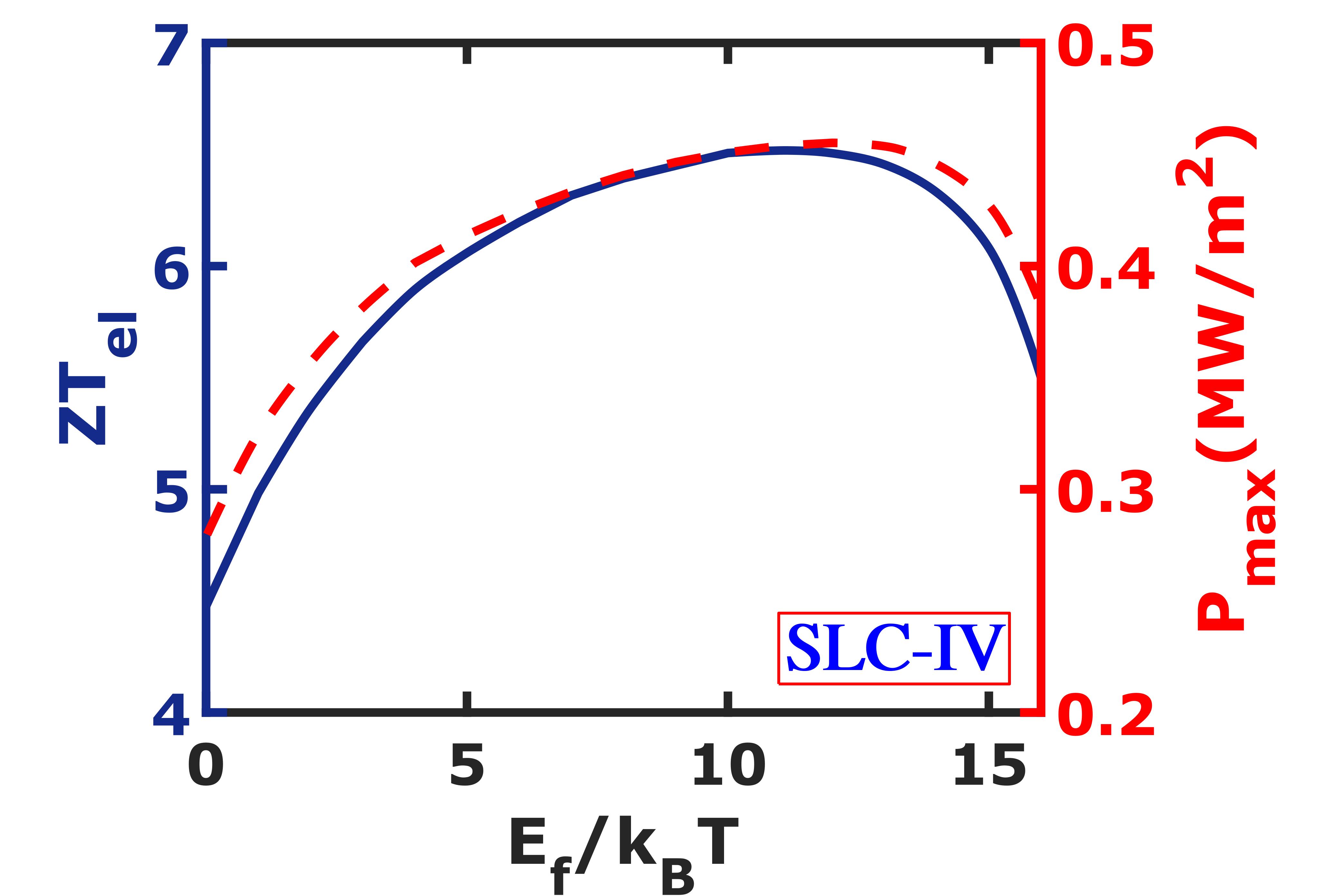}\label{11B_GW_ZT_Pmax_kT}}
	
	\caption{Figure of merit $zT_{el}$ and $P_{max}$ as a function of the electrochemical potential position for (a) SLC-I, (b) SLC-II, (c) SLC-III and (d) SLC-IV.}	
	\label{ZT_Pmax}
\end{figure}

\subsection{Figure-of-Merit Analysis}
We have so far restricted ourselves to the power-efficiency analysis, generally valid even under non-equilibrium conditions. However, in the linear response, it is customary to describe the TE performance via the dimensionless figure of merit $zT_{el}$ for electronic transport, which can be re-written as 
\begin{equation}
zT_{el} = \frac{S^2 G}{G_K} T_{avg},
\end{equation}
where $S$ is the Seebeck coefficient, $G$ is the electrical conductivity, $G_K$ is the electronic part of the thermal conductivity and $T_{avg}$ is the average of the hot ($T_H$) and cold ($T_C$) contact temperatures respectively. We extract these quantities using our simulation set up considering the coupled charge and heat current equations in the linear response regime \cite{Kim2011} given as
 \begin{equation}
I=G\Delta V + G_S\Delta T,\quad   I_Q = G_P\Delta V + G_Q\Delta T,
\label{eqI-IQ}
\end{equation}
where, $G$, $G_S$, $G_P$, $G_Q$ are related to the Onsager coefficients \cite{LNEDatta}, $\Delta V$ and $\Delta T$ are the applied bias and temperature gradients respectively. The Seebeck coefficient, given by $S=-G_S/G$, where $G$ is obtained when $\Delta T=0$ at a small bias voltage. Similarly, $G_S$ is obtained by setting $\Delta V=0$ at a finite $\Delta T$. Likewise, $G_P = I_Q/\Delta V$ when $\Delta T=0$ and $G_Q=I_Q/\Delta T$ when $\Delta V=0$ are obtained. The thermal conductivity is given as $G_K = G_Q - \frac{G_P G_S}{G}$.

In Fig.~\ref{ZT_Pmax}, we plot $zT_{el}$ and $P_{max}$ as a function of the electrochemical potential or Fermi level. We note an overall improvement in the figure of merit due to the use of SL structures \cite{Sofo-Mahan1994} and their associated miniband transmission features \cite{Hicks1992, Hicks1993, Hicks1996}. The plots for the SLC-(I-III) case, as seen in  Figs.~\ref{5B_NSL_ZT_Pmax_kT}-\ref{11B_GEc_ZT_Pmax_kT}, clearly point out that $zT_{el}$ is not a good predictor of electrical power performance. However, for SLC-IV we note a high $zT_{el}$ value of 6, which matches the point of maximum power obtained, and hence is a good indicator. We note that the performance discussed so far will be affected by phonon heat conduction also. However, the focus of our work was primarily to engineer the electronic part of heat conduction, keeping in mind that the phonon conduction can be minimized due to the presence of such nanostructured interfaces \cite{Minnich2009,Zebarjadi2012}. 
 
\section{Conclusion}
In summary, we have studied extensively the thermoelectric performance with respect to efficiency at maximum output power in various superlattice structures, with emphasis on self-consistent charging effect, solved using NEGF-Poisson formalism. Various possible configurations of superlattice heterostructures such as regular superlattices, anti-reflective and Gaussian distributed superlattices have been compared, and it was concluded that the superlattice system with a Gaussian distribution of barrier thickness offers the maximum transmitivity and hence the highest achievable efficiency at maximum output power. Furthermore, it is noted that the Gaussian distributed barrier thickness system is a good predictor of maximum power for a given figure of merit $zT$. We believe that the favorable thermoelectric transport properties predicted for these systems can attract considerable attention in the thermoelectrics community for the use of superlattices for power generation applications. With the existing advanced thin-film growth technology, the suggested superlattice structures can be achieved, and such optimized thermoelectric performances can be realized.

{\it{Acknowlegements:}} The authors acknowledge funding from Indian Space Research Organization as a part of the RESPOND grant.

\bibliographystyle{apsrev}

\bibliography{Reference}

\end{document}